\documentclass[twocolumn,showpacs,preprintnumbers,amsmath,amssymb]{revtex4-1}

\usepackage{graphicx,latexsym}
\usepackage{amsmath,amssymb,amsfonts}
\usepackage{bm}
\usepackage{braket}
\usepackage{mathrsfs}

\newcommand{\up}{\uparrow}
\newcommand{\down}{\downarrow}

\newcommand{\ep}{\varepsilon}
\def\lesssim{\ \raise.3ex\hbox{$<$}\kern-0.8em\lower.7ex\hbox{$\sim$}\ }
\def\gesim{\ \raise.3ex\hbox{$>$}\kern-0.8em\lower.7ex\hbox{$\sim$}\ }

\begin{document}
\title{Stable non-equilibrium Fulde-Ferrell-Larkin-Ovchinnikov state in a spin-imbalanced driven-dissipative Fermi gas loaded on a three-dimensional cubic optical lattice}
\author{Taira Kawamura\email{tairakawa@keio.jp}$^1$, Daichi Kagamihara$^2$ and Yoji Ohashi$^1$}
\affiliation{$^1$Department of Physics, Keio University, 3-14-1 Hiyoshi, Kohoku-ku, Yokohama 223-8522, Japan}
\affiliation{$^2$Department of Physics, Kindai University, Higashi-Osaka, Osaka 577-8502, Japan}
\date{\today}
\begin{abstract}
We theoretically investigate a Fulde-Ferrell-Larkin-Ovchinnikov (FFLO) type superfluid phase transition in a driven-dissipative two-component Fermi gas. The system is assumed to be in the non-equilibrium steady state, which is tuned by adjusting the chemical potential difference between two reservoirs that are coupled with the system. Including pairing fluctuations by extending the strong-coupling theory developed in the thermal-equilibrium state by Nozi\`eres and Schmitt-Rink to this non-equilibrium case, we show that a non-equilibrium FFLO (NFFLO) phase transition can be realized {\it without} spin imbalance, under the conditions that (1) the two reservoirs imprint a two-edge structure on the momentum distribution of Fermi atoms, and (2) the system is loaded on a three-dimensional cubic optical lattice. While the two edges work like two Fermi surfaces with different sizes, the role of the optical lattice is to prevent the NFFLO long-range order from destruction by NFFLO pairing fluctuations. We also draw the non-equilibrium mean-field phase diagram in terms of the chemical potential difference between the two reservoirs, a fictitious magnetic field to tune the spin imbalance of the system, and the environmental temperature of the reservoirs, to clarify the relation between the NFFLO state and the ordinary thermal-equilibrium FFLO state discussed in spin-imbalanced Fermi gases.
\end{abstract}
\maketitle
\section{Introduction}
The realization of unconventional superfluid states is one of the most exciting challenges in the current stage of cold Fermi gas physics. At present, although various non-$s$-wave pairing states have been discovered in metallic superconductivity, as well as in liquid $^3$He, the simplest isotropic $s$-wave superfluid state has only been realized in $^{40}$K \cite{Jin2004} and $^6$Li \cite{Zwierlein2004,Kinast2004,Bartenstein2004} Fermi gases. Since the high tunability is an advantage of ultracold Fermi gases \cite{Chin2010,Barontini2013, Labouvie2015, Labouvie2016, Tomita2017, Chong2018}, once this challenge is achieved, one would be able to examine its various superfluid properties in a wide parameter region. Indeed, in $^{40}$K and $^6$Li superfluid Fermi gases, a tunable pairing interaction associated with a Feshbach resonance \cite{Chin2010} has enabled systematic studies about how the weak-coupling Bardeen-Cooper-Schrieffer (BCS) type superfluid discussed in metallic superconductivity continuously changes to the Bose-Einstein condensation (BEC) of tightly bound molecules, with increasing the $s$-wave interaction strength \cite{Levin2005,Zwerger2012,Strinati2018,Ohashi2020}.
\par
\begin{figure}[tb]
\centering
\includegraphics[width=7.8cm]{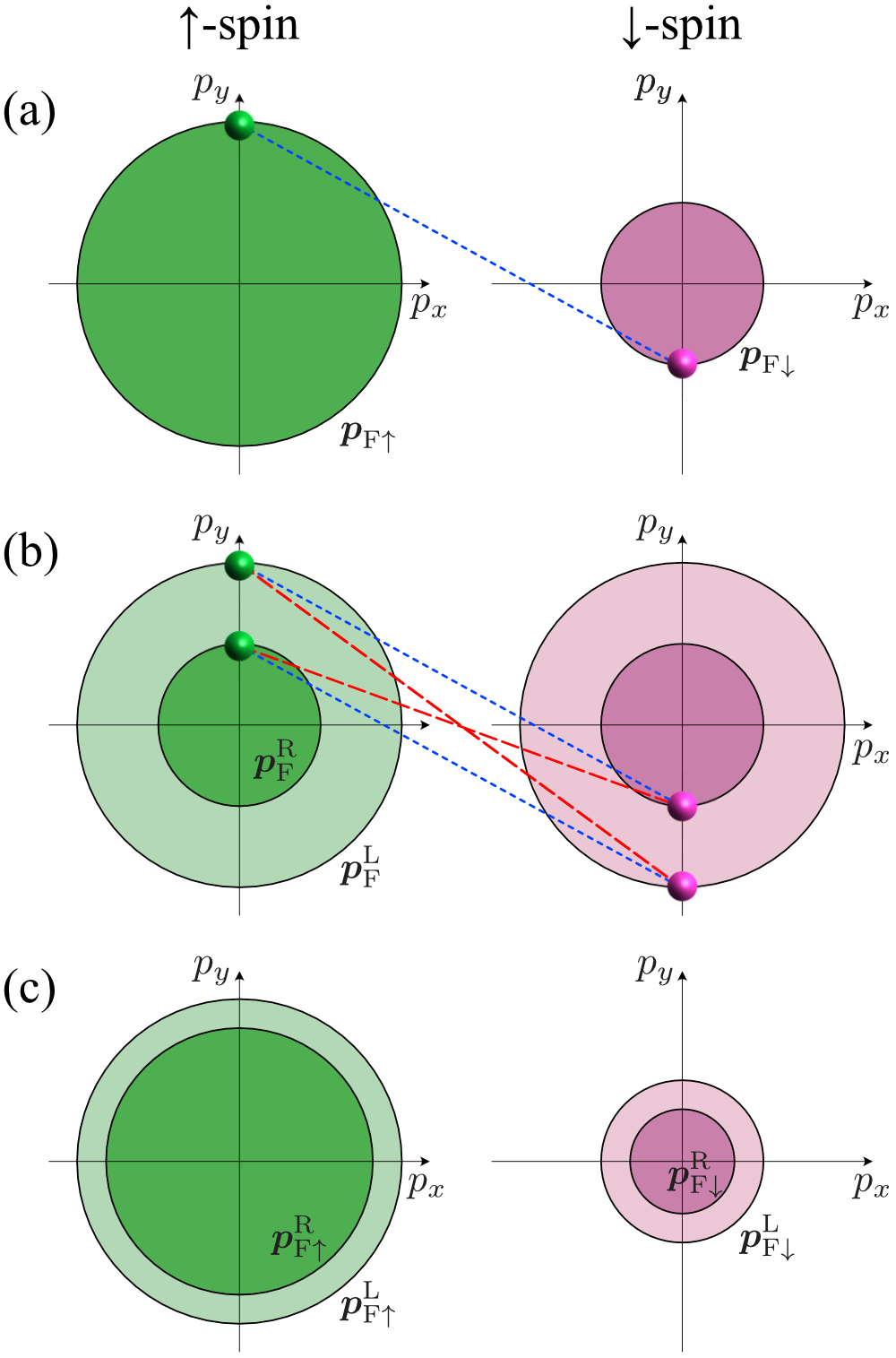}
\caption{(a) Cooper pair in the thermal-equilibrium FFLO state in a spin-imbalanced Fermi gas.  The large (small) colored circle with the radius $p_{{\rm F}\uparrow}$ ($p_{{\rm F}\downarrow}$) represents the Fermi surface of the $\uparrow$-spin ($\downarrow$-spin) component. The dotted line with two small circles at the ends denotes a FFLO Cooper pair. (b) NFFLO Cooper pairs. The color intensity schematically describes the particle occupancy in momentum space, and the edges at $p_{\rm F}^{\rm L}$ and $p_{\rm F}^{\rm R}$ (where the occupancy sharply changes) work like two Fermi surfaces with different radius. In this case, besides the conventional BCS-type Cooper pairs with zero center-of-mass momentum (dashed lines), the FFLO-type Cooper pairs (dotted lines) also become possible. In the latter case, Cooper pairs are formed between an $\uparrow$-spin Fermi atom near the edge at $p_{\rm F}^{\rm L}$ and a $\downarrow$-spin atom near the edge at $p_{\rm F}^{\rm R}$, as well as an $\uparrow$-spin atom near the edge at $p_{\rm F}^{\rm R}$ and an $\downarrow$-spin atom near the edge at $p_{\rm F}^{\rm L}$. (c) Particle occupation in the momentum space in a spin-{\it imbalanced} driven dissipative Fermi gas. When we simply call each edge `Fermi surface', this system looks as if there are four Fermi surfaces at $p_{{\rm F}\uparrow}^{\rm L}$, $p_{{\rm F}\uparrow}^{\rm R}$, $p_{{\rm F}\downarrow}^{\rm L}$, and $p_{{\rm F}\downarrow}^{\rm R}$.
}
\label{fig.1} 
\end{figure}

Among various candidates discussed in cold Fermi gas physics, the Fulde-Ferrell-Larkin-Ovchinnikov (FFLO) state \cite{Fulde1964, Larkin1964, Takada1969, Matsuda2007} is a promising one. This unconventional Fermi superfluid is characterized by a spatially oscillating superfluid order parameter $\Delta({\bm r})$, which is symbolically written as
\begin{equation}
\Delta({\bm r})=\Delta e^{i{\bm Q}_{\rm FF}\cdot{\bm r}},
\label{eq.0.1}
\end{equation}
where ${\bm Q}_{\rm FF}$ physically describes the center-of-mass-momentum of a FFLO Cooper pair. Although the FFLO state was originally proposed in the context of metallic superconductivity under an external magnetic field \cite{Fulde1964, Larkin1964, Takada1969, Matsuda2007}, it has also recently been discussed in ultracold Fermi gases \cite{Strinati2018, Hu2006, Parish2007, Liao2010, Chevy2010, Kinnunen2018}. Cooper pairs in the FFLO state are formed between Fermi surfaces with different sizes as shown in Fig.~\ref{fig.1}(a), giving ${\bm Q}_{\rm FF}\ne 0$. Recently, the observation of the FFLO state has been reported in several superconducting materials, such as CeCoIn$_5$ \cite{Bianchi2002,Bianchi2003,Kumagai2006}, CeCu$_2$Si$_2$ \cite{Kitagawa2018}, KFe$_2$As$_2$ \cite{Cho2017}, FeSe \cite{Ok2020,Kasahara2020}, and $\kappa$-(BEDT-TTF)$_2$Cu(NCS)$_2$ \cite{Singleton2000,Wright2011,Mayaffre2014}. Thus, the realization of a FFLO superfluid Fermi gas would be important, in order for cold atom physics to catch up with this recent exciting progress in condensed matter physics. 
\par
\par
At a glance, ultracold Fermi gases seem suitable for the FFLO state: (1) Although the FFLO state is known to be weak against impurity scatterings, one can prepare a very clean Fermi gas. (2) The splitting of Fermi surfaces between $\uparrow$-(pseudo)spin and $\downarrow$-(pseudo)spin components can immediately be prepared in a spin-{\it imbalanced} Fermi gas. 
\par
However, in spite of these advantages, the realization of the FFLO superfluid Fermi gas has not been reported yet. This seems because the current cold Fermi gas experiments are facing the following two serious difficulties: (i) In the presence of spin imbalance, the system undergoes the phase separation into the BCS superfluid region and the normal-state region of unpaired excess atoms, before reaching the desired FFLO phase transition \cite{Sheehy2006,He2008,Shin2008}. (ii) The spatial isotropy of the gas cloud anomalously enhances FFLO pairing fluctuations, which completely destroy the FFLO long-range order, even in three dimensions \cite{Shimahara1998, Ohashi2002, Radzihovsky2009, Radzihovsky2011, Yin2014, Jakubczyk2017, Wang2018, Zdybel2021, Kawamura2020_JLTP, Kawamura2020}.
\par
\begin{figure}[tb]
\centering
\includegraphics[width=8cm]{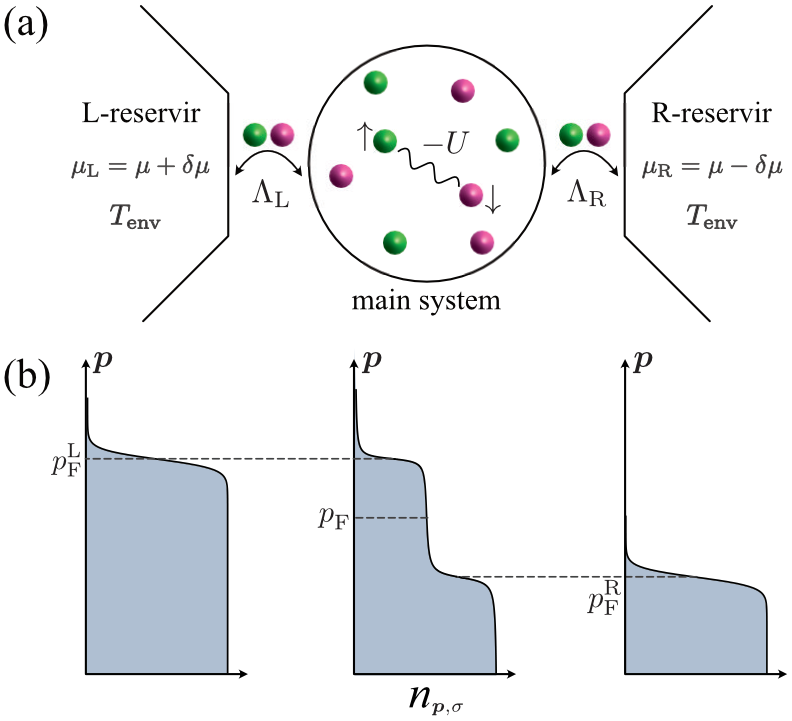}
\caption{(a) Model driven-dissipative two-component ($\sigma=\up,\down$) ultracold Fermi gas with a tunable $s$-wave pairing interaction $-U$ ($<0$). The central main system is coupled with two reservoirs ($\alpha={\rm L}, {\rm R}$) in the thermal equilibrium state, having different values of the Fermi chemical potentials, $\mu_{\rm L}=\mu+\delta\mu$ and $\mu_{\rm R}=\mu-\delta\mu$. Both reservoirs are free Fermi gases at the common environment temperature $T_{\rm env}$. $\Lambda_{\alpha}$ denotes a tunneling matrix element between the main system and the $\alpha$-reservoir. When $\delta\mu\ne 0$, the momentum distribution $n_{{\bm p},\sigma}$ of Fermi atoms has two edges around $p_{\rm F}^{\rm L}=\sqrt{2m\mu_{\rm L}}$ and $p_{\rm F}^{\rm R}=\sqrt{2m\mu_{\rm R}}$ at low temperatures (where $m$ is an atomic mass), as shown in panel (b). These edges correspond to the `Fermi surface' edges illustrated in Fig.~\ref{fig.1}(b).}
\label{fig.2} 
\end{figure}

Regarding these problems, we have recently proposed the following two ideas \cite{Kawamura2020_JLTP, Kawamura2020,Kawamura2022_2}: For (i), instead of using a spin-imbalanced Fermi gas, we proposed to use a spin-{\it balanced} driven dissipative Fermi gas, being coupled with two reservoirs as schematically shown in Fig.~\ref{fig.2}(a) \cite{Kawamura2020}. This is a non-equilibrium open system, where losses of particles and energy are continuously compensated by the two reservoirs \cite{Yamaguchi2012,Hanai2016,Hanai2017,Hanai2018}, and is known to exhibit various interesting phenomena that cannot be examined in the thermal equilibrium state \cite{Cross1993, Hoyle2006, Cross2009}. In this non-equilibrium system, we showed that, under a certain condition, the momentum distribution of Fermi atoms has a two-edge structure, originating from the chemical potential difference between the two reservoirs, as illustrated in Fig.~\ref{fig.2}(b). These edges work like two `Fermi surfaces' with different sizes, which enhances the FFLO pair correlation without spin imbalance. Indeed, we showed within the non-equilibrium mean-field theory that the FFLO superfluid state is obtained \cite{Kawamura2022}, where Cooper pairs are formed between Fermi atoms near the two edges, as schematically shown in Fig.~\ref{fig.1}(b). However, we also found that the difficulty (ii) also exists in this non-equilibrium case, so that this desired mean-field solution is immediately destroyed, once FFLO pairing fluctuations are taken into account \cite{Kawamura2020_JLTP, Kawamura2020}.
\par
For the difficulty (ii), we clarified in a {\it thermal-equilibrium} spin-imbalanced Fermi gas that the destruction of the FFLO long-range order by FFLO pairing fluctuations can be avoided, when the spatial isotropy of the gas cloud is removed by loading the system on a three-dimensional cubic optical lattice \cite{Kawamura2022_2}. However, we also found that, as far as we deal with a spin-{\it imbalanced} Fermi gas, the stabilized FFLO state competes with the above-mentioned phase separation, so that careful parameter tuning is still needed.
\par
\par
In this paper, by combining these two ideas, we explore a possible route to reach the FFLO superfluid phase transition in ultracold Fermi gases, {\it without} facing the phase separation, as well as the destruction by FFLO pairing fluctuations. For this purpose, we again consider the model driven-dissipative two-component Fermi gas shown in Fig.~\ref{fig.2}(a), but this time we impose a three-dimensional optical lattice potential to the system. To include pairing fluctuations, we extend the strong-coupling theory developed in the thermal-equilibrium state by Nozi\`eres and Schmitt-Rink (NSR) \cite{Nozieres1985}, to the case when the system is out of equilibrium. Using this, we examine whether or not the combined two-step structure of the Fermi momentum distribution with the background optical lattice can stabilize the non-equilibrium FFLO (NFFLO) state, overwhelming the above-mentioned difficulties.
\par
We note that the NFFLO state discussed in this paper is, strictly speaking, somehow different from the ordinary thermal-equilibrium FFLO state in the spin-imbalanced system: As illustrated in Fig.~\ref{fig.1}(a), the ordinary FFLO Cooper pairs are formed between $\uparrow$-spin particles near a larger Fermi surface and $\downarrow$-spin particles near a smaller one. In the NFFLO state, on the other hand, since the momentum distribution of each spin component has two edges [see Fig.~\ref{fig.1}(b)], Cooper pairs are formed between, not only $\uparrow$-spin particles near a larger Fermi surface edge and $\downarrow$-spin particles near a smaller Fermi surface edge, but also $\uparrow$-spin particles near the smaller Fermi surface edge and $\downarrow$-spin particles near the larger Fermi surface edge. 
\par
We also examine how these two kinds of FFLO states are related to each other, by considering the case with spin imbalance. In the thermal-equilibrium state, the spin imbalance causes the so-called Zeeman splitting between the $\uparrow$-spin and $\downarrow$-spin Fermi surfaces. When the system becomes out of equilibrium by adjusting the chemical potential difference between the two reservoirs, the spin imbalance further splits each edge structure in the momentum distribution into two, so that the system looks as if there are four Fermi surfaces [see Fig.~\ref{fig.1}(c)]. Including these `multiple Fermi surfaces' within the framework of the non-equilibrium mean-field BCS theory, we draw a superfluid phase diagram with respect to the environmental temperature, the chemical potential difference, and a fictitious magnetic field to tune the spin imbalance. 
\par
Before ending this section, we quickly summarize recent progress in the driven-dissipative system. Recent experimental progress has enabled us to examine, not only classical, but also quantum many-body driven-dissipative systems, such as exciton-polaritons \cite{Kasprzak2006, Carusotto2013, Sanvitto2016}, superconducting circuits \cite{Houck2012, Ma2019}, optical cavities \cite{Brennecke2013, Vaidya2018}, as well as ultracold atomic gases \cite{Barontini2013, Labouvie2015, Labouvie2016, Tomita2017, Chong2018}. At present, the same driven-dissipative ultracold Fermi gas as the model shown in Fig.~\ref{fig.2}(a) has not been realized, the realization would be possible within the current experimental technology, by extending the recent transport experiment on a $^6 {\rm Li}$ Fermi gas in a two-terminal configuration \cite{Brantut2012, Krinner2015, Husmann2015, Nagy2016, Krinner2017}, or employing a tilted triple-well optical trap \cite{Graefe2006, Mossmann2006, Liu2007}. For a more detailed experimental proposal, see Ref. \cite{Kawamura2022}.
\par
This paper is organized as follows. In Sec.~II, we explain how to extend the mean-field BCS theory, as well as the strong-coupling NSR theory, developed in the thermal equilibrium state to the non-equilibrium steady state of a driven-dissipative Fermi gas. In this extension, we take into account the effects of a background optical lattice, as well as spin imbalance. Using these theories, we examine the NFFLO phase transition and effects of pairing fluctuations in Sec.~III. We also show how the optical lattice stabilizes the NFFLO long-range order there. In Sec.~IV. we consider a driven-dissipative lattice Fermi gas with spin imbalance. In the phase diagram with respect to the environmental temperature, the chemical potential difference between the reservoirs, and a fictitious magnetic field to adjust the spin imbalance, we identify the region where the non-equilibrium BCS, NFFLO, and ordinary FFLO states appear. Throughout this paper, we set $\hbar=k_{\rm B}=1$, and the system volume $V$ is taken to be unity, for simplicity.
\par
\section{Formulation}
\par 
\subsection{Model driven-dissipative lattice Fermi gas}
\par
The model driven-dissipative non-equilibrium Fermi gas in Fig.~\ref{fig.2}(a) is described by the Hamiltonian,
\begin{equation}
H= H_{\rm sys} + H_{\rm env} +H_{\rm mix},
\label{eq.2.1}
\end{equation}
where
\begin{align}
H_{\rm sys}
&= 
\sum_{\bm{k}, \sigma} \ep_{\bm{k}} a^\dagger_{\bm{k},\sigma}a_{\bm{k}, \sigma} 
\notag\\
&\hspace{0.3cm}
-U\sum_{\bm{k}, \bm{k}', \bm{q}} a^\dagger_{\bm{k}+\bm{q}/2, \up} a^\dagger_{-\bm{k}+\bm{q}/2, \down}
a_{-\bm{k}'+\bm{q}/2, \down}a_{\bm{k}'+\bm{q}/2,\up},
\label{eq.Hsys}
\\
H_{\rm env} 
&=
\sum_{\alpha= {\rm L}, {\rm R}} \sum_{\bm{p}, \sigma}
\xi_{\bm{p}}^\alpha c^{\alpha \dagger}_{\bm{p},\sigma} c^{\alpha}_{\bm{p},\sigma},
\label{eq.Henv}
\\
H_{\rm mix} 
&=
\sum_{\alpha= {\rm L}, {\rm R}} \sum_{l,m=1}^N \sum_{\bm{p}, \bm{k}, \sigma} 
\Big[ e^{i\mu_{\alpha, \sigma} t} \Lambda_{\alpha} c^{\alpha \dagger}_{\bm{p},\sigma} a_{\bm{k}, \sigma} e^{-i\bm{p}\cdot\bm{R}_m^\alpha} e^{i\bm{k}\cdot \bm{r}_l} 
\notag\\
&\hspace{2.8cm}+{\rm H.c.} \Big].
\label{eq.Hmix}
\end{align}
Among the three, the attractive Hubbard Hamiltonian $H_{\rm sys}$ in Eq.~(\ref{eq.Hsys}) describes the main system, consisting of a two-component Fermi gas. This main system is loaded on a three-dimensional cubic optical lattice, in order to remove the spatial isotropy from the original gas system \cite{Kawamura2022_2}. $a_{\bm{k},\sigma}$ describes a Fermi atom with momentum ${\bm k}$ and pseudo-spin $\sigma=\uparrow,\downarrow$, which describe two atomic hyperfine states contributing to the Cooper-pair formation. The kinetic energy $\ep_{\bm{k}}$ of this lattice fermion has the form,
\begin{align}
\ep_{\bm{k}} 
=& 
-2t \sum_{j=x,y,z} \big[\cos(k_j)-1\big] \nonumber \\
&
-4t'\big[\cos(k_x)\cos(k_y) +\cos(k_y)\cos(k_z) 
\notag\\
&\hspace{3.3cm}
+\cos(k_z)\cos(k_x) -3\big],
\label{eq.dispersion}
\end{align}
where the lattice constant is taken to be unity, for simplicity. $t$ and $t'$ are the nearest-neighbor and the next-nearest-neighbor hopping parameters, respectively. The on-site $s$-wave pairing interaction $-U~(<0)$ in Eq.~(\ref{eq.Hsys}) is assumed to be tunable by adjusting the threshold energy of a Feshbach resonance \cite{Chin2010}.
\par
$H_{\rm env}$ in Eq.~(\ref{eq.Henv}) describes the left ($\alpha=L$) and right ($\alpha=R)$ reservoirs in Fig.~\ref{fig.2}(a). Here, $c^{\alpha}_{\bm{p},\sigma}$ is the annihilation operator of a Fermi atom in the $\alpha$-reservoir, with the kinetic energy $\xi^\alpha_{\bm{p}}=\bm{p}^2/(2m)-\omega_{\alpha}$ (where $m$ is an atomic mass). Each reservoir is assumed to be a huge free Fermi gas compared to the main system (which is satisfied by setting $\omega_{\alpha}$ to be much larger than the bandwidth of the main system) and is in the thermal equilibrium state at the common environmental temperature $T_{\rm env}$. Thus, the particle occupation in each reservoir obeys the ordinary Fermi distribution function,
\begin{equation}
f(\xi_{\bm k}^\alpha) = \frac{1}{e^{\xi_{\bm k}^\alpha/T_{\rm env}}+1}.
\end{equation}
We briefly note that, since the main system is out of equilibrium, the temperature is {\it not} well-defined there. Thus, in this paper, we use $T_{\rm env}$ as the temperature parameter in considering the superfluid phase transition of the non-equilibrium main system. 
\par
The system-reservoir coupling is described by the Hamiltonian $H_{\rm mix}$ in Eq.~(\ref{eq.Hmix}), where $\Lambda_{\alpha={\rm L}, {\rm R}}$ is a tunneling matrix element between the main system and the $\alpha$-reservoir. For simplicity, we set $\Lambda_{\rm L}=\Lambda_{\rm R}=\Lambda$ in the following discussions. In Eq.~(\ref{eq.Hmix}), the particle tunneling is assumed to occur between $i$th lattice-site at $\bm{r}_i$ and randomly distributing spatial positions $\bm{R}_j^\alpha$ in the $\alpha$ reservoir (where $i, j=1,\cdots, N_{\rm L} \gg 1$, with $N_{\rm L}$ being the number of lattice sites in the main system). Although the discrete translational symmetry associated with the background optical lattice is broken by these spatially random tunnelings, this symmetry property is recovered by taking the spatial average over the tunneling positions \cite{Kawamura2020, Kawamura2022, Hanai2016, Hanai2017, Hanai2018}. As discussed in Ref. \cite{Kawamura2022}, recent two-terminal experiments in cold atom physics \cite{Brantut2012, Krinner2015, Husmann2015, Nagy2016, Krinner2017} may effectively be regarded as spatially uniform systems. The random tunnelings assumed in our model are a theoretical trick to mimic such tunneling processes \cite{Kawamura2020, Kawamura2022}.
\par
\begin{figure}[tb]
\centering
\includegraphics[width=8.5cm]{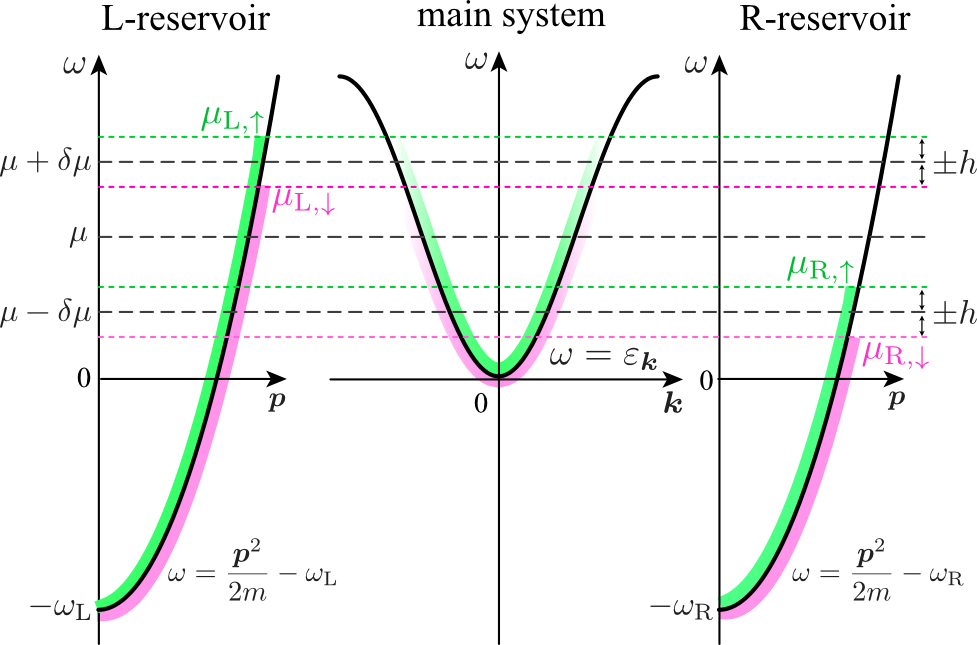}
\caption{Energy band in the main system, as well as those in the left and right reservoirs. The energy is commonly measured from the bottom ($\ep_{\bm{k}}=0$) of the band in the main system.  In the $\alpha$ reservoir at $T_{\rm env}=0$, the $\sigma$-spin band is filled up to the Fermi chemical potential $\mu_{\alpha,\sigma}$, given in Eqs.~(\ref{eq.del.muL}) and (\ref{eq.del.muR}).
}
\label{fig.3} 
\end{figure}

In Eq.~(\ref{eq.Hmix}), the factor $\exp\big(i\mu_{\alpha, \sigma} t\big)$ is introduced in order to describe the situation that the energy band of the $\sigma$-spin component in the $\alpha$ reservoir is filled up to the Fermi chemical potential $\mu_{\alpha,\sigma}$ at $T_{\rm env}=0$ \cite{Kawamura2020, Kawamura2022} (see Fig.~\ref{fig.3}). Thus, when we set $\mu_{\alpha,\uparrow}\ne \mu_{\alpha,\downarrow}$, the main system is in the spin-imbalanced state. For later convenience, we write $\mu_{\alpha,\sigma}$ as
\begin{align}
& \mu_{{\rm L},\sigma} = \mu +\sigma h +\delta\mu \equiv\mu_\sigma +\delta\mu,
\label{eq.del.muL}
\\[2pt]
& \mu_{{\rm R},\sigma} = \mu +\sigma h -\delta\mu \equiv\mu_\sigma -\delta\mu.
\label{eq.del.muR}
\end{align}
Here, $h$ is a fictitious magnetic field to tune the spin imbalance. $\delta\mu$ is half the chemical potential difference between the two reservoirs, which makes the main system out of equilibrium. In particular, this paper focuses on the non-equilibrium steady state, where the magnitude of the tunneling current from the left reservoir to the main system is equal to that from the main system to the right reservoir. We also impose the condition that the main system has no net current. (Of course, Fermi atoms flow across the junctions between the reservoirs and the main system.)
\par
\par
\subsection{Non-equilibrium mean-field (NMF) theory}
\par
\begin{figure}[tb]
\centering
\includegraphics[width=8.5cm]{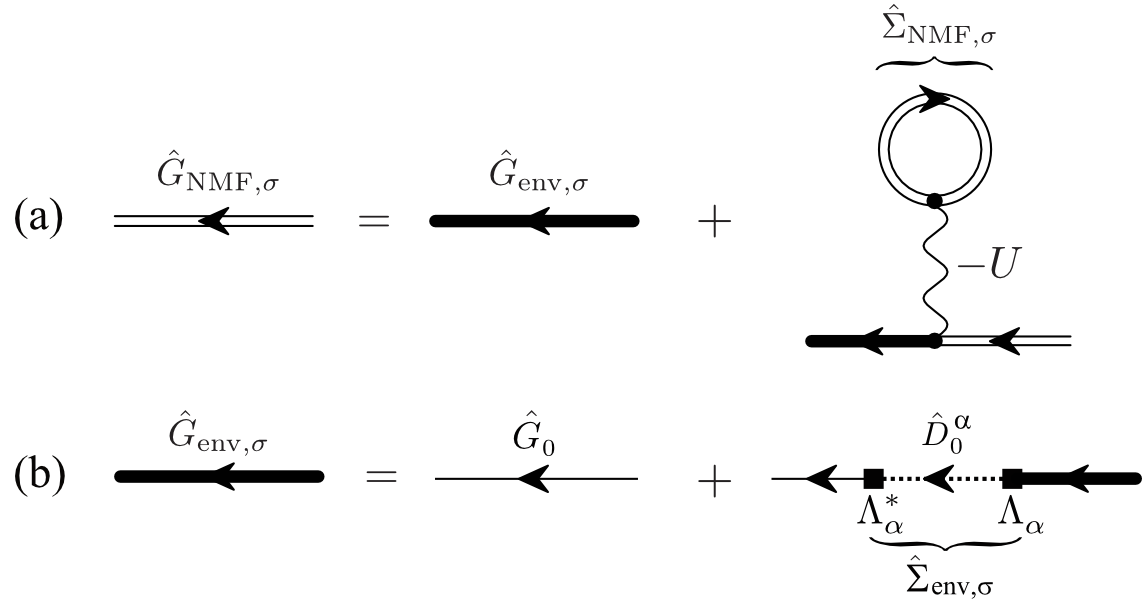}
\caption{(a) Dyson equation for the $2\times 2$ matrix single-particle non-equilibrium Green's function $\hat{G}_{{\rm NMF}, \sigma}$ (double solid line) in the main system. The self-energy $\hat{\Sigma}_{{\rm NMF}, \sigma}$ describes effects of the pairing interaction $-U$ (wavy line) within the NMF level. The thick solid line denotes $\hat{G}_{{\rm env},\sigma}$, which obeys the other Dyson equation in panel (b). The self-energy $\hat{\Sigma}_{{\rm env}, \sigma}$ involves effects of system-reservoir couplings within the second-order Born approximation. In panel (b), Green's functions ${\hat G}_{0,\sigma}$ and ${\hat D}_{0,\sigma}^\alpha$, respectively, describe free lattice fermions in the main system and a free Fermi gas in the $\alpha$-reservoir.}
\label{fig.4} 
\end{figure}

We first deal with the model Hamiltonian in Eq.~(\ref{eq.2.1}) within the non-equilibrium mean-field (NMF) level. Effects of pairing fluctuations are separately discussed in Sec.~II.C. To construct the NMF theory, we conveniently introduce the $2\times 2$ matrix single-particle non-equilibrium Green's function \cite{Rammer1986, Rammer2007, Stefanucci2013, Zagoskin2014},
\begin{equation}
\hat{G}_{{\rm NMF}, \sigma}(\bm{k}, \omega)=
\begin{pmatrix}
G^{\rm R}_{{\rm NMF},\sigma}(\bm{k}, \omega) & 
G^{\rm A}_{{\rm NMF},\sigma}(\bm{k}, \omega) \\[4pt]
0 & G^{\rm K}_{{\rm NMF},\sigma}(\bm{k}, \omega)
\end{pmatrix},
\label{eq.GNMF}
\end{equation}
where the superscripts `R', `A', and `K' represent the retarded, advanced, and Keldysh components, respectively. This NMF Green's function obeys the Dyson equation, which is diagrammatically described as Fig.~\ref{fig.4}(a). The expression for this equation is given by
\begin{align}
&\hat{G}_{{\rm NMF}, \sigma}(\bm{k}, \omega)
=
\hat{G}_{{\rm env}, \sigma}(\bm{k}, \omega) 
\notag\\[2pt]
&\hspace{1cm}+
\hat{G}_{{\rm env}, \sigma}(\bm{k}, \omega) 
\hat{\Sigma}_{{\rm NMF}, \sigma}(\bm{k}, \omega) 
\hat{G}_{{\rm NMF}, \sigma}(\bm{k}, \omega).
\label{eq.Dyson.NMF}
\end{align}
In Eq.~(\ref{eq.Dyson.NMF}), the $2\times 2$ matrix self-energy $\hat{\Sigma}_{{\rm NMF}, \sigma}(\bm{k}, \omega)$ has the form,
\begin{align}
\hat{\Sigma}_{{\rm NMF}, \sigma}(\bm{k}, \omega)
&=
\left(
\begin{array}{cc}
\Sigma_{{\rm NMF}, \sigma}^{\rm R}(\bm{k}, \omega) &
\Sigma_{{\rm NMF}, \sigma}^{\rm K}(\bm{k}, \omega) \\[4pt]
0 &
\Sigma_{{\rm NMF}, \sigma}^{\rm A}(\bm{k}, \omega)
\end{array}
\right)
\notag\\
&=Un_{{\rm NMF},-\sigma}
{\hat \tau}_0.
\label{eq.self.NMF2.R}
\end{align}
Here, ${\hat \tau}_0$ is the $2\times 2$ unit matrix acting on the Keldysh space, and 
\begin{equation}
n_{{\rm NMF}, -\sigma} 
=-i\sum_{\bm{k}} \int_{-\infty}^\infty \frac{d\omega}{2\pi} 
G^{<}_{{\rm NMF},-\sigma} (\bm{k}, \omega)
\label{eq.n.NMF.def}
\end{equation}
is the filling fraction of Fermi atoms at each lattice site in the main system (where `$-\sigma$' means the opposite component to $\sigma$) \cite{Rammer2007, Stefanucci2013, Zagoskin2014}. In Eq.~(\ref{eq.n.NMF.def}), the lesser Green's function $G^{<}_{{\rm NMF}, \sigma}$ is related to $G^{\rm R,A,K}_{{\rm NMF}, \sigma}$ as
\begin{align}
G^{<}_{{\rm NMF}, \sigma}(\bm{k}, \omega) 
= 
\frac{1}{2} \Big[ &-G^{\rm R}_{{\rm NMF}, \sigma}(\bm{k}, \omega) +G^{\rm A}_{{\rm NMF}, \sigma}(\bm{k}, \omega) 
\notag\\
&+G^{\rm K}_{{\rm NMF}, \sigma}(\bm{k}, \omega)\Big].
\label{eq.GL.NMF}
\end{align}
\par
\par
The Green's function $\hat{G}_{{\rm env}, \sigma}(\bm{k}, \omega)$ in Eq.~(\ref{eq.Dyson.NMF}) involves effects of the system-reservoir couplings $\Lambda_{\alpha={\rm L,R}}~(=\Lambda)$, and obeys the other Dyson equation which is diagrammatically drawn as Fig.~\ref{fig.4}(b). Taking the spatial average over the tunneling positions $\bm{r}_l$ and $\bm{R}_m^{\alpha}$ to recover the discrete translational symmetry of the cubic optical lattice \cite{Kawamura2020, Kawamura2022}, one finds that the Dyson equation for $\hat{G}_{{\rm env}, \sigma}(\bm{k}, \omega)$ can be written as
\begin{align}
\hat{G}_{{\rm env}, \sigma}(\bm{k}, \omega)
&=
\hat{G}_0(\bm{k}, \omega) 
\notag\\[2pt]
&\hspace{0.05cm}+
\hat{G}_0(\bm{k}, \omega) 
\hat{\Sigma}_{{\rm env}, \sigma}(\bm{k}, \omega) 
\hat{G}_{{\rm env}, \sigma}(\bm{k}, \omega),
\label{eq.Dyson.env}
\end{align}
where the self-energy $\hat{\Sigma}_{{\rm env}, \sigma}(\bm{k}, \omega)$ describes effects of the system-reservoir couplings. Within the second-order Born approximation, we have \cite{Kawamura2020, Kawamura2022}
\begin{align}
&\hat{\Sigma}_{{\rm env}, \sigma}(\bm{k}, \omega)
\notag\\
&=
\begin{pmatrix}
-2i\gamma &
-4i\gamma \big[1 -f(\omega-\mu_{{\rm L}, \sigma}) -f(\omega-\mu_{{\rm R}, \sigma}) \big] \\[6pt]
0 &
2i\gamma 
\end{pmatrix}.
\label{eq.self.env}
\end{align}
Here, $\gamma=\pi N^2 |\Lambda|^2 \rho$ works as the quasi-particle damping rate due to the system-reservoir couplings, where $\rho$ is the single-particle density of states in the reservoirs. In obtaining Eq.~(\ref{eq.self.env}), we have ignored the $\alpha$ $(={\rm L}, {\rm R})$ dependence of the density of states $\rho$, for simplicity. We have also employed the so-called wide-band limit approximation \cite{Stefanucci2013}, which ignores the $\omega$ dependence of $\rho$.
\par
In the Dyson equation (\ref{eq.Dyson.env}), 
\begin{equation}
\hat{G}_{0}(\bm{k}, \omega)
=
\begin{pmatrix}
\frac{1}{\omega+i\delta -\ep_{\bm{k}}} &
-2\pi i \delta(\omega -\ep_{\bm{k}}) \big[1 -2f_{\rm ini}(\omega)\big] \\[6pt]
0 &
\frac{1}{\omega-i\delta -\ep_{\bm{k}}}
\end{pmatrix}
\label{eq.G0}
\end{equation}
is the bare Green's function in the assumed thermal equilibrium initial state at $t=-\infty$, where the system-reservoir couplings $\Lambda$, as well as the pairing interaction $-U$, were absent. In Eq.~(\ref{eq.G0}), $f_{\rm ini}(\omega)=1/[e^{\omega/T_{\rm ini}}+1]$ is the Fermi distribution function at the initial temperature $T_{\rm ini}$ in the main system, and $\delta$ is an infinitesimally small positive number. 
\par
We briefly note that the bare Green's function $\hat{D}_{0,\alpha}(\bm{k}, \omega)$ in the $\alpha$-reservoir, which appears in Fig.~\ref{fig.4}(b), has the same form as $\hat{G}_{0}(\bm{k}, \omega)$ in Eq.~(\ref{eq.G0}) where the single-particle energy $\ep_{\bm{k}}$ and the initial temperature $T_{\rm ini}$ are replaced by $\xi^\alpha_{\bm p}$ and $T_{\rm env}$, respectively.
\par
Solving the Dyson equation (\ref{eq.Dyson.env}), one obtains
\begin{equation}
\hat{G}_{{\rm env}, \sigma}(\bm{k}, \omega)=
\begin{pmatrix}
\frac{1}{\omega -\ep_{\bm{k}} +2i\gamma} &
\frac{-4i\gamma[1 -f(\omega -\mu_{{\rm L},\sigma}) -f(\omega -\mu_{{\rm R},\sigma})]}
{[\omega -\ep_{\bm{k}}]^2 +4\gamma^2}\\[6pt]
0 &
\frac{1}{\omega -\ep_{\bm{k}} -2i\gamma}
\end{pmatrix}.
\label{eq.Genv}
\end{equation}
Here, we emphasize that the Fermi distribution function $f(\omega -\mu_{\alpha,\sigma})$ in Eq.~(\ref{eq.Genv}) has nothing to do with the `initial' Fermi distribution function $f_{\rm ini}(\omega)$ appearing in the bare Green's function at $t=-\infty$ given in Eq.~(\ref{eq.G0}). This means that $\hat{G}_{{\rm env}, \sigma}$ no longer has the initial memory of the system at $t=-\infty$, which is wiped out by the couplings with the two reservoirs \cite{Stefanucci2013}.
\par
Substituting Eqs.~(\ref{eq.self.NMF2.R}) and (\ref{eq.Genv}) into the Dyson equation \eqref{eq.Dyson.NMF}, we have
\begin{align}
&
\hat{G}_{{\rm NMF}, \sigma}(\bm{k}, \omega) 
\notag\\
&\hspace{0.3cm}=
\begin{pmatrix}
\frac{1}{\omega -\tilde{\ep}_{\bm{k}, \sigma} +2i\gamma} &
\frac{-4i\gamma[1 -f(\omega -\mu_{{\rm L},\sigma}) -f(\omega -\mu_{{\rm R},\sigma})]}
{[\omega -\tilde{\ep}_{\bm{k},\sigma}]^2 +4\gamma^2}\\[6pt]
0 &
\frac{1}{\omega -\tilde{\ep}_{\bm{k}, \sigma} -2i\gamma}
\end{pmatrix},
\label{eq.GNMF2}
\end{align}
where the kinetic energy $\tilde{\ep}_{\bm{k}, \sigma}$ involves the Hartree energy as
\begin{equation}
\tilde{\ep}_{\bm{k}, \sigma} = \ep_{\bm{k}} -U n_{{\rm NMF}, -\sigma}. 
\label{eq.tild.ep}
\end{equation}
The expression for the filling fraction $n_{{\rm NMF}, \sigma}$ of Fermi atoms in the main system is obtained from Eqs.~(\ref{eq.n.NMF.def}), (\ref{eq.GL.NMF}) and (\ref{eq.GNMF2}), as
\begin{align}
n_{{\rm NMF}, \sigma} 
&= 
\sum_{\bm{k}} \int_{-\infty}^\infty \frac{d\omega}{2\pi} \frac{4\gamma\big[f(\omega -\mu_{{\rm L}, \sigma}) +f(\omega -\mu_{{\rm R}, \sigma})\big]}{[\omega -\tilde{\ep}_{\bm{k}, \sigma}]^2 +4\gamma^2}
\notag\\[4pt]
&\equiv
\sum_{\bm{k}} f^{\rm neq}_{\bm{k}, \sigma}.
\label{eq.n.NMF}
\end{align}
\par
To grasp how the momentum distribution $f^{\rm neq}_{\bm{k}, \sigma}$ in the main system is `engineered' by the two reservoirs ($\alpha={\rm L,R}$), it is convenient to take the small damping limit $\gamma\to 0$, giving
\begin{equation}
f^{\rm neq}_{\bm{k},\sigma}|_{\gamma\to+0}
=
\frac{1}{2} \big[f(\tilde{\ep}_{\bm{k}, \sigma} -\mu_{\sigma}-\delta\mu) +f(\tilde{\ep}_{\bm{k}, \sigma} -\mu_{\sigma} +\delta\mu)\big].
\label{eq.small_gamma}
\end{equation}
Equation (\ref{eq.small_gamma}) is found to exhibit two edges around the momenta $\bm{k}^{\rm R}_{{\rm F}\sigma}$ and $\bm{k}^{\rm L}_{{\rm F}\sigma}$, which satisfy, respectively,
\begin{align}
&\tilde{\ep}_{\bm{k}^{\rm R}_{{\rm F}\sigma}, \sigma} = \mu_{\sigma} -\delta \mu 
\label{eq.eff.kR}
,\\[4pt]
&\tilde{\ep}_{\bm{k}^{\rm L}_{{\rm F}\sigma}, \sigma} = \mu_{\sigma} +\delta \mu
\label{eq.eff.kL}.
\end{align}
As mentioned previously, these edges function like two `Fermi surfaces' with different sizes \cite{Kawamura2022}.
\par
\begin{figure}[tb]
\centering
\includegraphics[width=8.5cm]{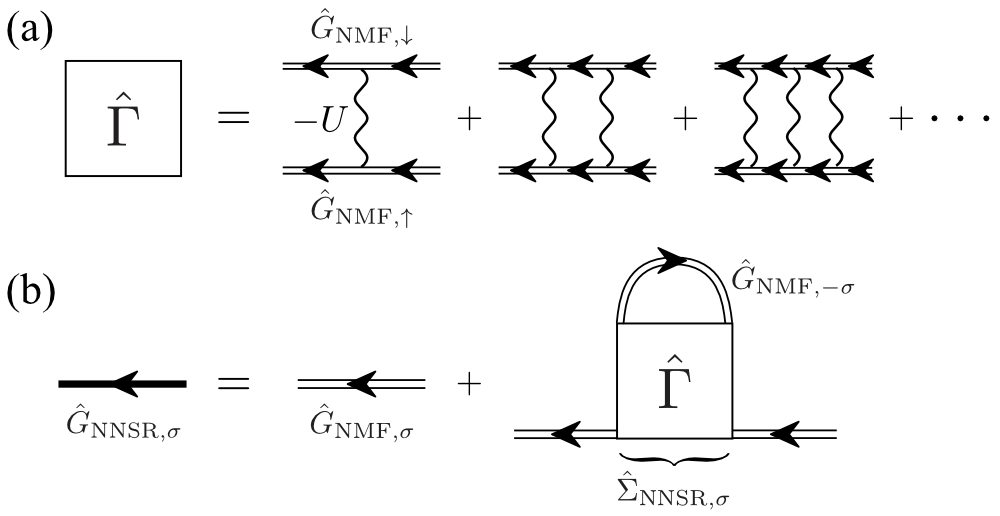}
\caption{(a) Non-equilibrium $2\times 2$ particle-particle scattering matrix $\hat{\Gamma}$ in Keldysh space. The double solid line is $\hat{G}_{{\rm NMF}, \sigma}$ given in Eq.~(\ref{eq.GNMF2}). (b) Truncated Dyson equation giving the NNSR single-particle Green's function $\hat{G}_{{\rm NNSR}, \sigma}$ (thick solid line) in the main system. The self-energy $\hat{\Sigma}_{{\rm NNSR}, \sigma}$ describes effects of pairing fluctuations.
}
\label{fig.5} 
\end{figure}
\par
In the NMF scheme, we obtain the environmental temperature $T_{\rm env}^{\rm c}$ at which the main system experiences the superfluid phase transition. For this purpose, we extend the theory developed by Kadanoff and Martin (KM) \cite{Kadanoff1961, KadanoffBook} to the present model. In the KM theory, the key is the pole ($\equiv \nu_{\bm{q}}$) of the retarded particle-particle scattering matrix $\Gamma^{\rm R}(\bm{q},\nu)$: In the normal phase ($T_{\rm env}> T_{\rm env}^{\rm c}$), $\Gamma^{\rm R}(\bm{q},\nu)$ has a complex pole in the lower-half complex plain (${\rm Im}[\nu_{\bm{q}}]<0$), which physically means the stability of the system, because pairing fluctuations decay in time. When $T_{\rm env} <T_{\rm env}^{\rm c}$, on the other hand, a pole appears in the {\it upper} half plain (${\rm Im}[\nu_{\bm{q}}]>0$) (see Fig.~4 in Ref.~\cite{Kawamura2020}), indicating the instability of the system against pairing fluctuations that exponentially grow in time. Thus, the superfluid phase transition is determined from the `KM condition' that $\Gamma^{\rm R}(\bm{q},\nu)$ has a {\it real} pole \cite{Kadanoff1961, KadanoffBook, Kawamura2020}. 
\par
\begin{widetext}
The particle-particle scattering matrix $\hat{\Gamma}(\bm{q}, \nu)$ in the NMF theory is described by the ladder-type diagrams shown in Fig.~\ref{fig.5}(a) \cite{Kawamura2020}. Summing up these diagrams, one has \cite{Kawamura2020}
\begin{align}
\hat{\Gamma}(\bm{q}, \nu)=
\begin{pmatrix}
\Gamma^{\rm R}(\bm{q},\nu) &  
\Gamma^{\rm K}(\bm{q},\nu) \\[4pt]
0 & 
\Gamma^{\rm A}(\bm{q},\nu) 
\end{pmatrix}
&=
\begin{pmatrix}
\frac{-U}{1+U\Pi^{\rm R}(\bm{q}, \nu)} & 
\frac{U^2 \Pi^{\rm K}(\bm{q}, \nu)}{[1+U\Pi^{\rm R}(\bm{q}, \nu)] [1+U\Pi^{\rm A}(\bm{q}, \nu)]}
\\[6pt]
0 &
\frac{-U}{1+U\Pi^{\rm A}(\bm{q}, \nu)}
\end{pmatrix},
\label{eq.Tmat}
\end{align}
where
\begin{align}
\Pi^{\rm R}(\bm{q}, \nu)
&=
\big[\Pi^{\rm A}(\bm{q}, \nu)\big]^*
\notag\\
&=
\frac{i}{2} \sum_{\bm{k}} \int_{-\infty}^\infty \frac{d\omega}{2\pi}
\Big[G^{\rm R}_{{\rm NMF}, \up}(\bm{k}+\bm{q}/2, \omega +\nu) G^{\rm K}_{{\rm NMF}, \down}(-\bm{k}+\bm{q}/2, -\omega) 
\notag\\
&\hspace{6cm}+
G^{\rm K}_{{\rm NMF}, \up}(\bm{k}+\bm{q}/2, \omega +\nu) G^{\rm R}_{{\rm NMF}, \down}(-\bm{k}+\bm{q}/2, -\omega)\Big]
\label{eq.PiR},\\
\Pi^{\rm K}(\bm{q}, \nu)
&=
\frac{i}{2} \sum_{\bm{k}} \int_{-\infty}^\infty \frac{d\omega}{2\pi}
\Big[G^{\rm R}_{{\rm NMF}, \up}(\bm{k}+\bm{q}/2, \omega +\nu) G^{\rm R}_{{\rm NMF}, \down}(-\bm{k}+\bm{q}/2, -\omega) 
\notag\\
&\hspace{6cm}+
G^{\rm A}_{{\rm NMF}, \up}(\bm{k}+\bm{q}/2, \omega +\nu) G^{\rm A}_{{\rm NMF}, \down}(-\bm{k}+\bm{q}/2, -\omega)
\notag\\
&\hspace{6cm}+
G^{\rm K}_{{\rm NMF}, \up}(\bm{k}+\bm{q}/2, \omega +\nu) G^{\rm K}_{{\rm NMF}, \down}(-\bm{k}+\bm{q}/2, -\omega)\Big],
\label{eq.PiK}
\end{align}
are the pair correlation functions. From retarded component of Eq.~(\ref{eq.Tmat}), the pole of $\Gamma^{\rm R}(\bm{q}, \nu)$ is obtained by solving
\begin{equation}
1 + U\Pi^{\rm R}(\bm{q}, \nu_{\bm{q}})= 0.
\label{eq.pole}
\end{equation}
Since $\Pi^{\rm R}(\bm{q}, \nu)$ in Eq.~\eqref{eq.PiR} is a complex function, the pole equation \eqref{eq.pole} actually consists of two equations, that is, $1 + U{\rm Re}\big[\Pi^{\rm R}(\bm{q}, \nu_{\bm{q}})\big]= 0$ and ${\rm Im}[\Pi^{\rm R}(\bm{q}, \nu_{\bm{q}})]= 0$. Between the two, the latter reads
\begin{equation}
0= 
\sum_{\alpha={\rm L}, {\rm R}} \sum_{\bm{k}} \int_{-\infty}^\infty \frac{d\omega}{2\pi}
\frac{
\tanh\left(\frac{\omega+\nu_{\bm{q}}/2 -\mu_{\alpha, \up}}{2T^{\rm c}_{\rm env}}\right) +
\tanh\left(\frac{-\omega+\nu_{\bm{q}}/2 -\mu_{\alpha, \down}}{2T^{\rm c}_{\rm env}}\right)}
{\big[(\omega -\nu_{\bm{q}}/2 +\tilde{\ep}_{\bm{k}+\bm{q}/2, \up})^2 +4\gamma^2 \big]
 \big[(\omega +\nu_{\bm{q}}/2 -\tilde{\ep}_{-\bm{k}+\bm{q}/2, \down})^2 +4\gamma^2 \big]}.
\label{eq.pole.Im}
\end{equation}
One finds that Eq.~\eqref{eq.pole.Im} is satisfied only when $\nu_{\bm{q}}=2\mu$. Substituting this into the real part of the pole equation (\ref{eq.pole}), we obtain the $T_{\rm env}^{\rm c}$-equation,
\begin{equation}
1=U\gamma\sum_{\bm{k}} \int_{-\infty}^\infty \frac{d\omega}{2\pi}
\frac
{\big[2\omega + \tilde{\ep}_{\bm{k}+\bm{q}/2, \up} -\tilde{\ep}_{-\bm{k}+\bm{q}/2, \down} -2h\big]
\left[\tanh\left(\frac{\omega-\delta\mu}{2T^{\rm c}_{\rm env}}\right) +\tanh\left(\frac{\omega-\delta\mu}{2T^{\rm c}_{\rm env}}\right)\right]}
{\big[(\omega +\tilde{\ep}_{\bm{k}+\bm{q}/2, \up} -\mu_\up)^2 +4\gamma^2 \big]
 \big[(\omega -\tilde{\ep}_{\bm{k}+\bm{q}/2, \down} +\mu_\down)^2 +4\gamma^2 \big]}.
\label{eq.NThouless}
\end{equation}
In the NMF theory, one solves the $T_{\rm env}^{\rm c}$-equation~(\ref{eq.NThouless}), together with the equation for the filling fraction in Eq.~(\ref{eq.n.NMF}), to self-consistently determine $T_{\rm env}^{\rm c}$ and $\mu(T_{\rm env}^{\rm c})$ for a given parameter set $(\delta\mu,n_\uparrow,n_\downarrow)$. In the $T_{\rm env}^{\rm c}$-equation~(\ref{eq.NThouless}), the momentum $\bm{q}$ is chosen so as to obtain the highest $T_{\rm env}^{\rm c}$. The self-consistent solution with $\bm{Q}_{\rm FF}\neq 0$ ($\bm{Q}_{\rm FF}$ is the optimal value of ${\bm q}$) describes the superfluid phase transition into the spatially non-uniform NFFLO state, where each Cooper pair has the non-zero center-of-mass momentum $\bm{Q}_{\rm FF}$. On the other hand, the uniform solution with $\bm{Q}_{\rm FF}=0$ means the non-equilibrium BCS (NBCS) state. 

\par
\par
\subsection{Non-equilibrium Nozi\`eres-Schmitt-Rink (NNSR) theory}
\par
We now include the effects of pairing fluctuations by extending the strong coupling theory developed in the thermal equilibrium state by Nozi\`{e}res-Schmitt-Rink (NSR) \cite{Nozieres1985} to the case when the system is out of equilibrium. In this non-equilibrium NSR (NNSR) scheme, the single-particle Green's function in the main system is given by
\begin{align}
\hat{G}_{{\rm NNSR}, \sigma}(\bm{k}, \omega)
&=
\begin{pmatrix}
G^{\rm R}_{{\rm NNSR},\sigma}(\bm{k}, \omega) & 
G^{\rm A}_{{\rm NNSR},\sigma}(\bm{k}, \omega) \\[4pt]
0 & G^{\rm K}_{{\rm NNSR},\sigma}(\bm{k}, \omega) 
\end{pmatrix}
\nonumber\\[6pt]
&=
\hat{G}_{{\rm NMF}, \sigma}(\bm{k}, \omega) 
+
\hat{G}_{{\rm NMF}, \sigma}(\bm{k}, \omega) 
\hat{\Sigma}_{{\rm NNSR}, \sigma}(\bm{k}, \omega)
\hat{G}_{{\rm NMF}, \sigma}(\bm{k}, \omega),
\label{eq.Gsys}
\end{align}
where $\hat{G}_{{\rm NMF}, \sigma}$ is given by Eq.~(\ref{eq.GNMF2}). Here, the NNSR self-energy,
\begin{equation}
\hat{\Sigma}_{{\rm NNSR}, \sigma}(\bm{k}, \omega)=
\begin{pmatrix}
\Sigma^{\rm R}_{{\rm NNSR},\sigma}(\bm{k}, \omega) & 
\Sigma^{\rm A}_{{\rm NNSR},\sigma}(\bm{k}, \omega) \\[4pt]
0 & \Sigma^{\rm K}_{{\rm NNSR},\sigma}(\bm{k}, \omega) 
\end{pmatrix},
\end{equation}
is obtained from the evaluation of the second diagram in Fig.~\ref{fig.5}(b), which gives
\begin{align}
\Sigma^{\rm R}_{{\rm NNSR},\sigma}(\bm{k}, \omega) &=
\big[\Sigma^{\rm A}_{{\rm NNSR},\sigma}(\bm{k}, \omega)\big]^*
\notag\\[4pt]
&=
-\frac{i}{2} \sum_{\bm{q}} \int_{-\infty}^\infty \frac{d\nu}{2\pi}\Big[
\Gamma^{\rm R}(\bm{q}, \nu) G^{\rm K}_{{\rm NMF}, -\sigma}(\bm{q}-\bm{k}, \nu -\omega) +
\Gamma^{\rm K}(\bm{q}, \nu) G^{\rm A}_{{\rm NMF}, -\sigma}(\bm{q}-\bm{k}, \nu -\omega)
\Big]
\label{eq.selfR.NNSR}
,\\
\Sigma^{\rm K}_{{\rm NNSR},\sigma}(\bm{k}, \omega) &=
-\frac{i}{2} \sum_{\bm{q}} \int_{-\infty}^\infty \frac{d\nu}{2\pi}\Big[
\Gamma^{\rm A}(\bm{q}, \nu) G^{\rm R}_{{\rm NMF}, -\sigma}(\bm{q}-\bm{k}, \nu -\omega)
\notag\\
&\hspace{2.6cm}+
\Gamma^{\rm R}(\bm{q}, \nu) G^{\rm A}_{{\rm NMF}, -\sigma}(\bm{q}-\bm{k}, \nu -\omega) +
\Gamma^{\rm K}(\bm{q}, \nu) G^{\rm K}_{{\rm NMF}, -\sigma}(\bm{q}-\bm{k}, \nu -\omega)
\Big],
\label{eq.selfK.NNSR}
\end{align}
where the particle-particle scattering matrices $\Gamma^{\rm R, A, K}(\bm{q}, \nu)$ are given in Eq.~(\ref{eq.Tmat}). 
\par
As usual, the Fermi filling fraction $n_\sigma$ in the main system is obtained from the Keldysh component $G^{\rm K}_{{\rm NNSR},\sigma}(\bm{k}, \omega)$ in Eq.~(\ref{eq.Gsys}):
\begin{align}
n_\sigma 
&= \frac{i}{2} \sum_{\bm{k}} \int_{-\infty}^\infty \frac{d\omega}{2\pi} 
G^{\rm K}_{{\rm NNSR}, \sigma}(\bm{k}, \omega) -\frac{1}{2}
\notag\\
&= n_{{\rm NMF}, \sigma} + \frac{i}{2} 
\sum_{\bm{k}} \int_{-\infty}^\infty \frac{d\omega}{2\pi}
\Big[ 
\hat{G}_{{\rm NMF}, \sigma}(\bm{k}, \omega) 
\hat{\Sigma}_{{\rm NNSR}, \sigma}(\bm{k}, \omega)
\hat{G}_{{\rm NMF}, \sigma}(\bm{k}, \omega)
\Big]^{\rm K}
\notag\\
&\equiv
n_{{\rm NMF}, \sigma}  + n_{{\rm FL}, \sigma}.
\label{eq.filling.NNSR}
\end{align}
Here, $n_{{\rm NMF}, \sigma}$ is given in Eq.~\eqref{eq.n.NMF}, $n_{{\rm FL}, \sigma}$ is the NNSR strong-coupling correction to the filling fraction, and
\begin{equation}
\big[\hat{A}\hat{B}\hat{C}\big]^{\rm K} =
A^{\rm R} B^{\rm R} C^{\rm K} + A^{\rm R} B^{\rm K} C^{\rm A} +A^{\rm K} B^{\rm A} C^{\rm A}.
\end{equation}
\end{widetext}
\par
As the ordinary (thermal equilibrium) NSR theory \cite{Nozieres1985}, the NNSR theory also solves the $T_{\rm env}^{\rm c}$-equation (\ref{eq.NThouless}) that the NMF theory uses, together with Eq.~(\ref{eq.filling.NNSR}), to self-consistently determine the superfluid phase transition $T_{\rm env}^{\rm c}$, $\mu(T_{\rm env}^{\rm c})$, and ${\bm Q}_{\rm FF}$.
\par
Here, we comment on the values of parameters in our numerical calculations: (1) For the filling fraction $n_\sigma$, we set $n_\sigma<0.5$. This is because, although fluctuations in the particle-hole channel are known to become strong near the half-filling ($n_\sigma=0.5$) \cite{Tamaki2008}, the NNSR theory only includes fluctuations in the Cooper channel. (2) For the interaction strength $-U$, we deal with the weak-coupling regime, because the (N)FFLO state requires sharp Fermi edges. (3) The damping rate is chosen so as to satisfy $\gamma/(6t)\gesim 0.005$, due to computational problems, the reason for which is explained in Appendix~A. 
\par
\section{Stabilization of the NFFLO state in spin-balanced driven-dissipative lattice Fermi gas}
\par
In this section, we deal with the spin-{\it balanced} case, by setting $h=0$ in Eqs.~(\ref{eq.del.muL}) and (\ref{eq.del.muR}). The spin-imbalanced case is separately discussed in Sec.~IV.
\par
\begin{figure}[tb]
\centering
\includegraphics[width=8.5cm]{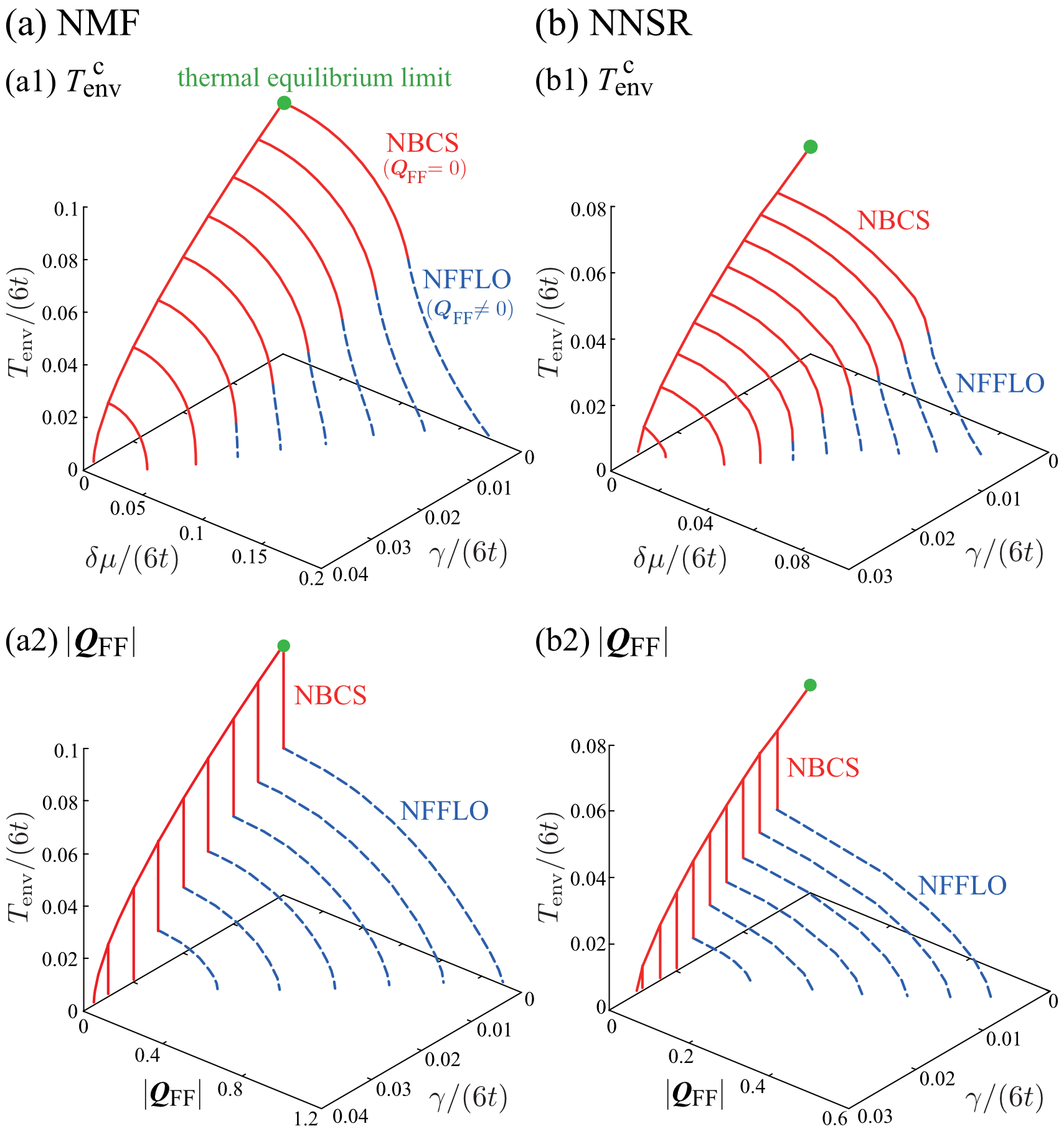}
\caption{Calculated $T_{\rm env}^{\rm c}$ (upper panels) and $|{\bm Q}_{\rm FF}|$ (lower panels) in a driven-dissipative {\it spin-balanced} lattice Fermi gas, as functions of the chemical potential bias $\delta\mu$ and the damping rate $\gamma$. (a) NMF theory. (b) NNSR theory. The solid (dashed line) line is the phase boundary between the normal state and the NBCS (NFFLO) state with $|{\bm Q}_{\rm FF}|=0$ ($|{\bm Q}_{\rm FF}|>0$). We take $t'=0$, $n=n_\up+n_\down =0.3$, and $U/(6t)=0.8$. (Note that $6t$ is the bandwidth in the main system, when $t'=0$.) The thermal equilibrium limit is at $\delta\mu=0$ and $\gamma\to+0$.}
\label{fig.6} 
\end{figure}
\begin{figure}[tb]
\centering
\includegraphics[width=8.5cm]{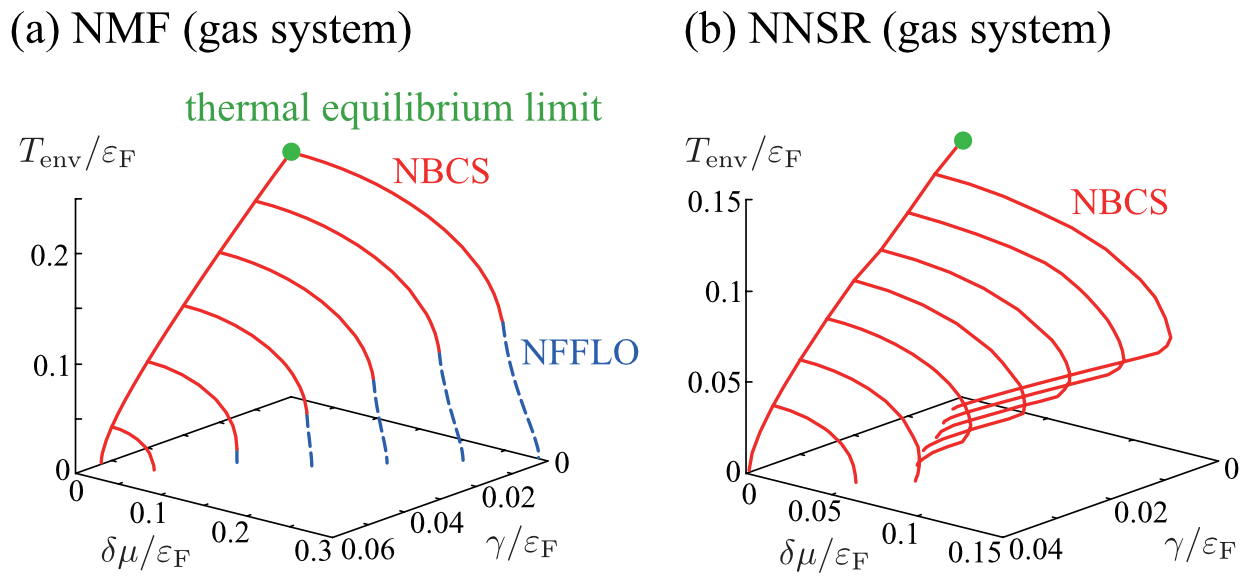}
\caption{Same plots as the upper panels in Fig.~\ref{fig.6}, in the case when the optical lattice is absent. We set $(p_{\rm F}a_s)^{-1}=-0.6$, where $a_s$ is the $s$-wave scattering length. $p_{\rm F}$ and $\ep_{\rm F}$ as $\ep_{\rm F}=p_{\rm F}^2/(2m)$ are, respectively, the Fermi momentum and the Fermi energy of a free Fermi gas with the particle number $N=p_{\rm F}^3/(3\pi^2)$. In panel (b), $T_{\rm env}^{\rm c}$ is seen to exhibit re-entrant behavior, due to the complete destruction of the NFFLO long-range order by anomalously enhanced NFFLO pairing fluctuations \cite{Kawamura2020_JLTP, Kawamura2020}.}
\label{fig.7} 
\end{figure}

The upper panels in Fig.~\ref{fig.6} show $T_{\rm env}^{\rm c}$ in a driven-dissipative spin-{\it balanced} Fermi gas, loaded on the cubic optical lattice. In this figure, we distinguish between the NBCS and NFFLO phase transitions from whether $|{\bm Q}_{\rm FF}|$ equals zero or not in the lower panels in Fig.~\ref{fig.6}. In contrast to the case with no optical lattice (where the mean-field NFFLO solution is completely destroyed by NFFLO pairing fluctuations \cite{Kawamura2020_JLTP, Kawamura2020}, as shown in Fig.~\ref{fig.7}), Figs.~\ref{fig.6}(a1) and (b1) show that the NFFLO long-range order survives against pairing fluctuations in the lattice system. (We summarize the NMF and the NNSR theories in the absence of the optical lattice in Appendix~B.)
\par
\par
\par
\begin{figure}[b]
\centering
\includegraphics[width=7.5cm]{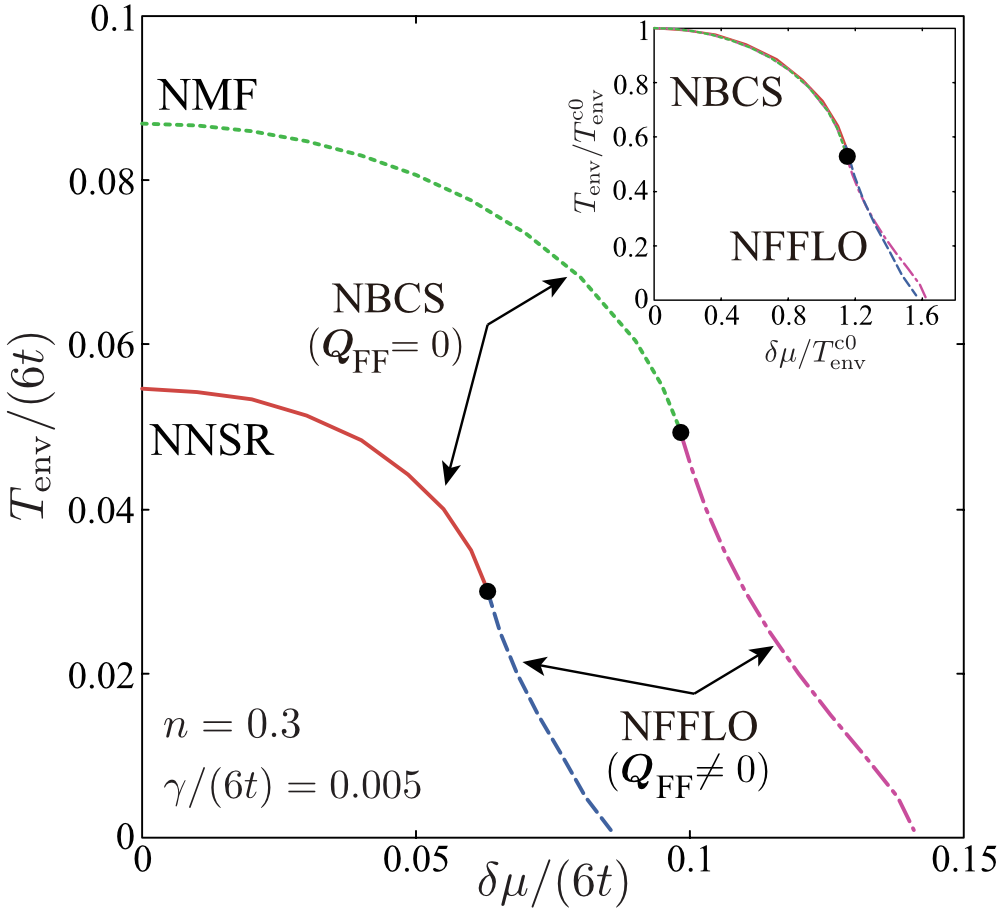}
\caption{Calculated $T_{\rm env}^{\rm c}$ as a function of $\delta\mu$. We set $t'=0$, $n=n_\up+n_\down =0.3$, $\gamma/(6t)=0.005$, and $U/(6t)=0.8$. In each NMF and NNSR result, the solid circle is the boundary between the NBCS and NFFLO phase transitions. The inset shows the results when $T_{\rm env}^{\rm c}$ and $\delta\mu$ are normalized by the superfluid phase transition temperature $T_{\rm env}^{\rm c0}$ at $\delta\mu=0$.}
\label{fig.8} 
\end{figure}
\par
We find from the comparison of Fig.~\ref{fig.6}(a1) with Fig.~\ref{fig.6}(b1) that pairing fluctuations tend to decrease $T_{\rm env}^{\rm c}$. However, apart from this difference, these figures also indicate that, once the NFFLO state is stabilized by the optical lattice, the NMF theory (which completely ignores pairing fluctuations) already captures the essential behavior of $T_{\rm env}^{\rm c}$ as a function of $\delta\mu$ and $\gamma$, at least when the filling fraction equals $n=n_\uparrow+n_\downarrow=0.3$. Indeed, as an example, when we extract $T_{\rm env}^{\rm c}$ at $\gamma/(6t)=0.005$ from Figs.~\ref{fig.6}(a1) and (b1), the NNSR result is found to be very similar to the NMF result, as shown in Fig.~\ref{fig.8}. Furthermore, when Fig.~\ref{fig.8} is replotted with respect to the scaled variables $T_{\rm env}^{\rm c}/T_{\rm env}^{\rm c0}$ and $\delta\mu/T_{\rm env}^{\rm c0}$ (where $T_{\rm env}^{\rm c0}$ is the superfluid phase transition temperature at $\delta\mu=0$), the scaled NMF and NFFLO results almost coincide with each other, as shown in the inset in Fig.~\ref{fig.8}. That is, although pairing fluctuations remarkably damage the mean-field NFFLO solution in the absence of optical lattice, their effects are not so crucial in a lattice Fermi gas, at least when $n=0.3$.
\par
\begin{figure}[tb]
\centering
\includegraphics[width=8.5cm]{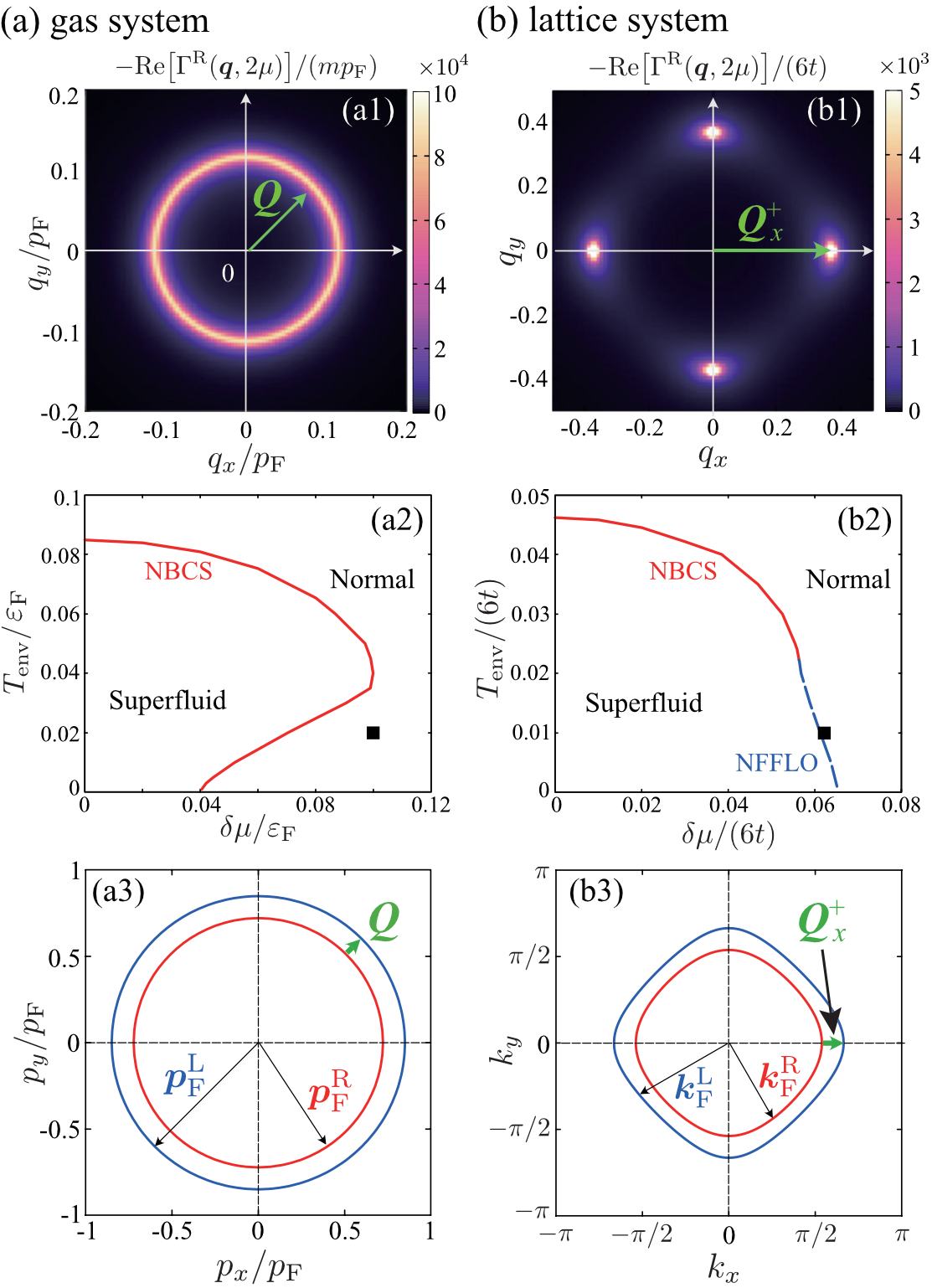}
\caption{(a1) Calculated intensity $-{\rm Re}\big[\Gamma^{\rm R}(\bm{q}=(q_x,q_y,0), \nu=2\mu)\big]$ of the real part of the retarded particle-particle scattering at the solid square in (a2), in the absence of optical lattice. (a3) Positions of two edges imprinted on the momentum distribution of Fermi atoms by the two reservoirs, at $|\bm{p}_{\rm F}^{\alpha}|=\sqrt{2m\mu_\alpha}=\sqrt{2m[\mu\pm \delta\mu]}$. $|{\bm Q}|$ in (a1) is just related to the size difference between the two edge circles shown in (a3). In calculating (a1)-(a3), we set $(p_{\rm F} a_s)^{-1}=-0.6$, and $\gamma/\ep_{\rm F}=0.02$. (b1)-(b3) show the case in the presence of the three-dimensional optical lattice: (b1) is the same plot as (a1) at the solid square in (b2). (b3) is the same plot as (a3), determined from Eqs.~(\ref{eq.eff.kR}) and (\ref{eq.eff.kL}). ${\bm Q}^+_x$ in (b1) is related to the size difference between the two Fermi surface edges shown in (b3). We set $U/(6t)=0.8$, $\gamma/(6t)=0.01$, $t'=0$, and $n=0.3$ in (b1)-(b3). 
}
\label{fig.9}
\end{figure}
\par
Figure~\ref{fig.9}(a1) shows the intensity $-{\rm Re}\big[\Gamma^{\rm R}(\bm{q}, \nu=2\mu)\big]$ of the retarded particle-particle scattering matrix near the region where the re-entrant phenomenon of the NBCS phase transition occurs [solid square in Fig.~\ref{fig.9}(a2)], in the absence of the optical lattice. Since this quantity physically describes pairing fluctuations with the center-of-mass momentum $\bm{q}$, the fact of the large intensity around $|{\bm q}|=|{\bm Q}|\ne 0$ means the enhancement of NFFLO pairing fluctuations there. We also point out that $|{\bm Q}|$ is directly related to the size difference between two `Fermi surface' edges that are imprinted on the momentum distribution of Fermi atoms by the two reservoirs [see Fig.~\ref{fig.9}(a3)]. This means that these edges really work like two Fermi surfaces with different sizes, as in the spin-imbalanced case.
\par
Figure~\ref{fig.9}(b1) shows the results in the presence of the three-dimensional cubic optical lattice, which is quite different from Fig.~\ref{fig.9}(a1). [Figure~\ref{fig.9}(b1) is obtained at the solid square in Fig.~\ref{fig.9}(b2).]: Since the spatial isotropy is broken by the optical lattice, the ring structure seen in Fig.~\ref{fig.9}(a1) is not obtained, but the $-{\rm Re}\big[\Gamma^{\rm R}(\bm{q}, \nu=2\mu)\big]$ exhibits four peaks, reflecting the four-fold rotational symmetry of the cubic lattice. However, for example, the peak at ${\bm q}={\bm Q}^+_x$ in Fig.~\ref{fig.9}(b1) is still related to the size difference between the two Fermi surface edges, as shown in Fig.~\ref{fig.9}(b3). That is, these edges also work like two Fermi surfaces, to enhance NFFLO pairing fluctuations around ${\bm q}={\bm Q}^+_x$, as well as the other equivalent peaks in Fig.~\ref{fig.9}(b1).
\par
The above-mentioned difference seen in Figs.~\ref{fig.9}(a1) and (b1) makes a significant difference in the NNSR fluctuation correction terms in Eqs.~(\ref{eq.filling.NNSR}) and (\ref{eq.filling.NNSR.gas}): In the presence of the optical lattice, noting that the particle-particle scattering matrix $\hat{\Gamma}(\bm{q}, \nu)$ in Eq.~\eqref{eq.Tmat} is enhanced around $(\bm{q}, \nu)=(\bm{Q}^\eta_j, 2\mu)$ near the NFFLO phase transition (where ${\bm Q}^{\eta=\pm}_{j=x,y,z}$ represent the four peak positions in Fig.~\ref{fig.9}(b1), as well as other two peak positions existing along the $q_z$ axis), we approximate the self-energy in Eqs.~\eqref{eq.selfR.NNSR} and \eqref{eq.selfK.NNSR} to, near $T_{\rm env}^{\rm c}$,
\begin{align}
&\hat{\Sigma}_{{\rm NNSR},\sigma}(\bm{k}, \omega)	
\notag\\
& \simeq
\scalebox{0.92}{$\displaystyle
-\Delta_{\rm pg}^2
\sum_{\eta=\pm} \sum_{j=x,y,z}
\begin{pmatrix}
G^{\rm A}_{{\rm NMF}, -\sigma} & 
G^{\rm K}_{{\rm NMF}, -\sigma} \\[4pt]
0 & 
G^{\rm R}_{{\rm NMF}, -\sigma}
\end{pmatrix}
(\bm{Q}^\eta_j-\bm{k}, 2\mu -\omega) $}
\notag\\[4pt]
&=
\Delta_{\rm pg}^2 \sum_{\eta=\pm} \sum_{j=x,y,z} \hat{G}^*_{{\rm NMF}, -\sigma}(\bm{Q}^\eta_j-\bm{k}, 2\mu -\omega).
\label{eq.PG.self}
\end{align}
Here, the so-called pseudogap parameter,
\begin{equation}
\Delta_{\rm pg}^2 = \frac{i}{2} \sum_{\bm{q}} \int_{-\infty}^\infty \frac{d\nu}{2\pi} \Gamma^{\rm K}(\bm{q}, \nu),
\label{eq.Delta.PG}
\end{equation}
physically describes the strength of pairing fluctuations \cite{Chen2005, Tsuchiya2009}. Evaluating the fluctuation correction $n_{{\rm FL}, \sigma}$ involved in Eq.~(\ref{eq.filling.NNSR}) by using Eq.~(\ref{eq.PG.self}), one has
\begin{align}
n_{{\rm FL}, \sigma}
&=
\frac{i\Delta_{\rm pg}^2}{2}
\sum_{\eta=\pm} \sum_{j=x,y,z} \sum_{\bm{k}} \int_{-\infty}^\infty \frac{d\omega}{2\pi}
\Big[ 
\hat{G}_{{\rm NMF}, \sigma}(\bm{k}, \omega) 
\notag\\
&\hspace{0.5cm}\times
\hat{G}^*_{{\rm NMF}, -\sigma}(\bm{Q}_j^\eta-\bm{k}, 2\mu -\omega)
\hat{G}_{{\rm NMF}, \sigma}(\bm{k}, \omega)
\Big]^{\rm K}.
\label{eq.app.NFL.lattice}
\end{align}
To evaluate the pseudogap parameter $\Delta_{\rm pg}^2$, we approximate $\Gamma^{\rm R}(\bm{q}, \nu)$ to
\begin{equation}
\Gamma^{\rm R}(\bm{q}, \nu) \simeq \sum_{\eta=\pm} \sum_{j=x,y,z}
\frac{-U}{C\big[\bm{q} -\bm{Q}_j^\eta \big]^2 -i\lambda \big[\nu -2\mu\big]},
\label{eq.app.TR.lattice}
\end{equation}
where we have assumed that $T_{\rm env}^{\rm c}$ satisfies Eq.~(\ref{eq.NThouless}), and 
\begin{align}
& C = \frac{U}{2} \left. \nabla_{\bm{q}}^2  \Pi^{\rm R}(\bm{q}, 2\mu) \right|_{\bm{q}=\bm{Q}_j^\pm},\\[4pt]
& \lambda= \frac{\pi U}{8T_{\rm env}} N(\mu) {\rm sech}^2\left(\frac{\delta\mu}{2T_{\rm env}}\right),
\label{eq.lambda}
\end{align}
with $N(\mu)$ being the density of states in the main system at $\omega=\mu$. In obtaining Eq.~(\ref{eq.app.TR.lattice}), for simplicity, we have taken the limit $\gamma\to+0$ in $\Pi^{\rm R}(\bm{q}, \nu)$ in Eq.~(\ref{eq.PiR}), giving
\begin{align}
&\lim_{\gamma\to +0}\Pi^{\rm R}(\bm{q}, \nu)
\notag\\
&=
\sum_{\bm{p}} \frac{1 -f(\tilde{\ep}_{\bm{p}+\bm{q}/2, \up} -\mu-\delta\mu)-f(\tilde{\ep}_{\bm{p}-\bm{q}/2, \down} -\mu+\delta\mu)} {\nu_+ -\tilde{\ep}_{\bm{p}+\bm{q}/2,\up} -\tilde{\ep}_{\bm{p}-\bm{q}/2, \down}},
\end{align}
and have expanded it around $(\bm{q}, \nu)=(\bm{Q}_{j=x,y,z}^{\eta=\pm}, 2\mu)$. Using Eqs.~\eqref{eq.Tmat} and \eqref{eq.app.TR.lattice}, one reaches
\begin{align}
\Delta_{\rm pg}^2 
&=
\frac{i}{2} \sum_{\bm{q}} \int_{-\infty}^\infty \frac{d\nu}{2\pi} 
\left|\Gamma^{\rm R}(\bm{q}, \nu)\right|^2 \Pi^{\rm K}(\bm{q}, \nu)
\notag\\
&\simeq
\sum_{\eta=\pm} \sum_{j=x,y,z} \frac{iU^2\Pi^{\rm K}(\bm{Q}_j^\eta, 2\mu)}{2} 
\notag\\
&\hspace{0.8cm}\times
\sum_{\bm{q}} \int_{-\infty}^\infty \frac{d\nu}{2\pi} \frac{1}{C^2 \big[\bm{q} -\bm{Q}_j^\eta\big]^4 +\lambda^2 \big[\nu-2\mu\big]^2}
\notag\\[6pt]
&=
\sum_{\eta=\pm} \sum_{j=x,y,z} \frac{iU^2\Pi^{\rm K}(\bm{Q}_j^\eta, 2\mu)}{4 \lambda C} 
\sum_{\bm{q}} \frac{1}{\big[\bm{q} -\bm{Q}_j^\eta\big]^2}.
\label{eq.PG.lattice}
\end{align}
In deriving the second line, we have employed the same approximation as that used in deriving Eq.~(\ref{eq.PG.self}). Replacing $\bm{q} -\bm{Q}_j^\eta$ by $\bm{q}$ in Eq.~\eqref{eq.PG.lattice}, one finds that $\Delta_{\rm pg}^2$ converges in three dimension, irrespective of the value of $\bm{Q}_j^\eta$. This immediately concludes the convergence of $n_{{\rm FL}, \sigma}$ [which is proportional to $\Delta_{\rm pg}^2$, see Eq.~\eqref{eq.app.NFL.lattice}]. Thus, the NNSR coupled equations (\ref{eq.NThouless}) and (\ref{eq.filling.NNSR}) can be satisfied simultaneously at the NFFLO phase transition (where ${\bm Q}_j^\eta\ne 0$). We briefly note that the six NFFLO vectors ${\bm Q}_{j=x,y,z}^{\eta=\pm}$ have the same magnitude, being equal to $|{\bm Q}_{\rm FF}|$ in Fig.~\ref{fig.6}(b2).
\par
\begin{figure}[tb]
\centering
\includegraphics[width=8cm]{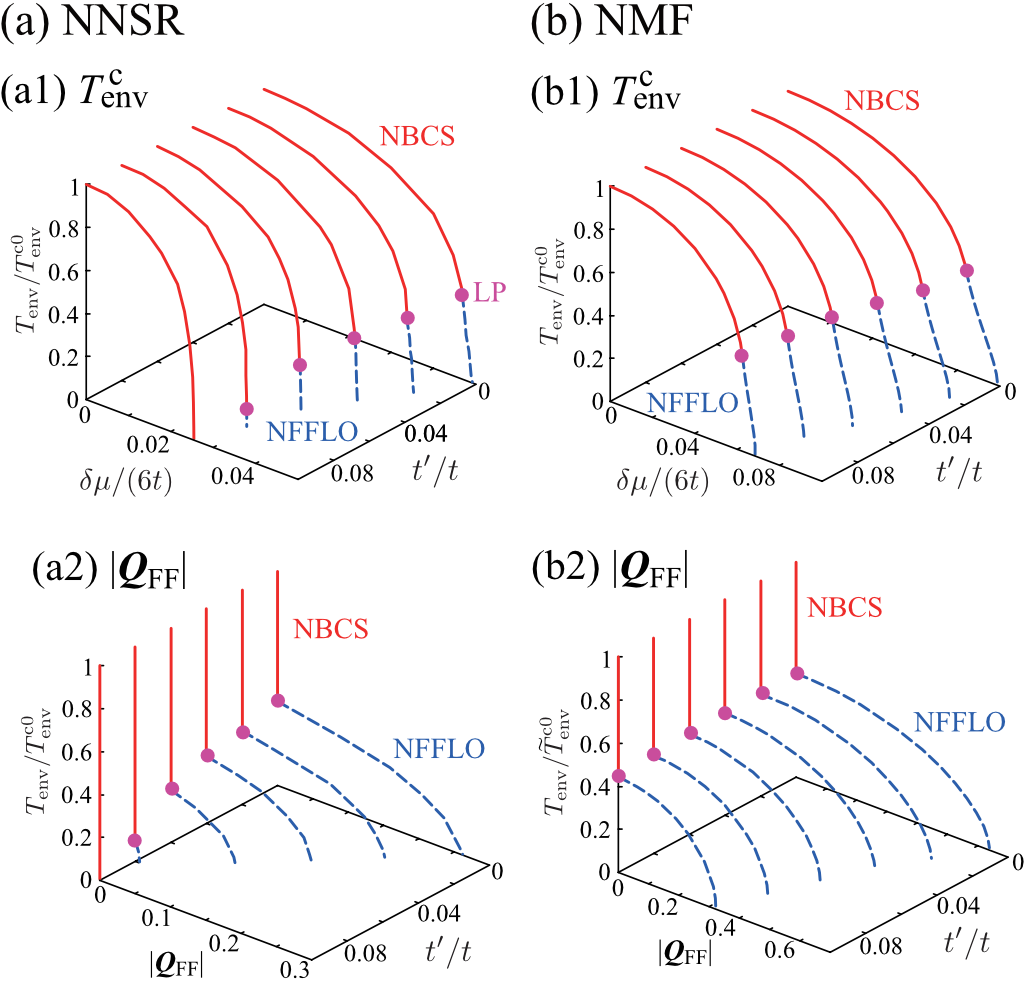}
\caption{Calculated $T_{\rm env}^{\rm c}$ (upper panels) and $|{\bm Q}_{\rm FF}|$ (lower panels) in a driven-dissipative ultracold lattice Fermi gas, as functions of the chemical potential bias $\delta\mu$ and the next nearest neighbor hopping $t'$. (a) NMF theory. (b) NNSR theory. The solid circle is the boundary between the NBCS (solid line) and NFFLO (dashed line) phase transitions, which is also referred to as the Lifshitz point in the literature. In the NMF case, the temperature $T^{\rm LP}_{\rm env}$ at the Lifshitz point is always located at $T^{\rm LP}_{\rm env}/T^{{\rm c}0}_{\rm env}\simeq 0.45$, irrespective of the value of $t'$ (at least within the parameter region shown in this figure). In contrast, $T^{\rm LP}_{\rm env}$ decreases with increasing $t'$ in the NNSR case. We set $U/(6t)=0.8$, $\gamma/(6t)=0.015$ and $n=0.3$. $T_{\rm env}^{\rm c}$ is normalized by the value at $\delta\mu=0$ ($\equiv T_{\rm env}^{{\rm c}0}$).}
\label{fig.10}
\end{figure}
\par
\begin{figure}[bt]
\centering
\includegraphics[width=7cm]{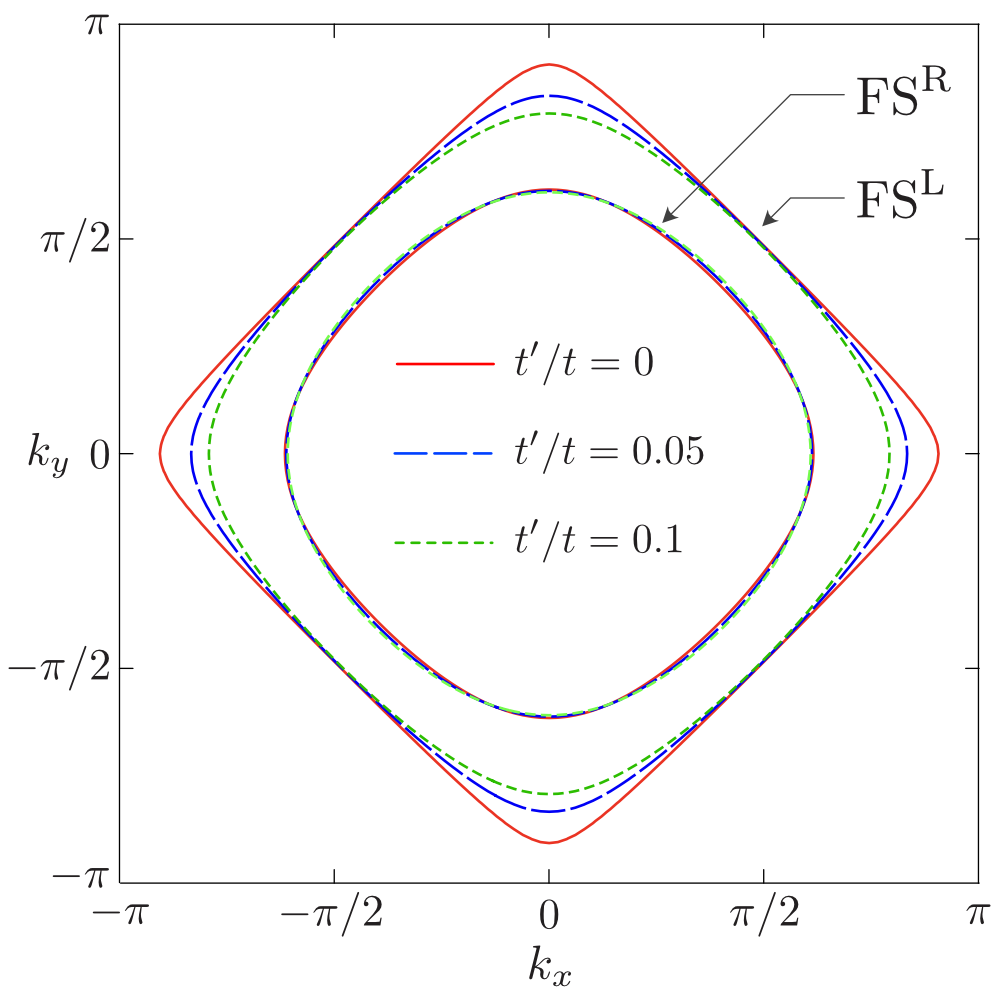}
\caption{Positions of `Fermi surface' edges (`FS$^{\rm L}$' and `FS$^{\rm R}$') produced by the two reservoirs for various values of the next nearest-neighbor hopping $t'$. We take $U=0$, $n=0.3$, $\delta\mu/(6t)=0.1$, $\gamma\to +0$, and $k_z=0$.}
\label{fig.11}
\end{figure}
\par
A quite different phenomenon occurs in the absence of the optical lattice: In this spatially isotropic case, when the intensity $-{\rm Re}\big[\Gamma^{\rm R}(\bm{q}, \nu=2\mu)\big]$ of the retarded particle-particle scattering matrix exhibits a ring structure as seen in Fig.~\ref{fig.9}(a1), the pseudogap parameter $\Delta_{\rm pg}^2$ can be approximated to
\begin{equation}
\Delta_{\rm pg}^2 \simeq \frac{iU^2\Pi^{\rm K}(\bm{Q}, 2\mu)}{4 \lambda C} \int_0^{q_{\rm c}}\frac{q^2 dq}{2\pi^2} \frac{1}{\big[|\bm{q}| -|\bm{Q}|\big]^2},
\label{eq.PG.gas}
\end{equation}
where $q_{\rm c}$ is a cutoff momentum. (For the derivation, see Appendix C.) Comparing Eq.~(\ref{eq.PG.gas}) with Eq.~(\ref{eq.PG.lattice}), the factor $1/\big[\bm{q} -\bm{Q}_j^\eta\big]^2$ in the lattice case is now replaced by $1/\big[|\bm{q}|-|\bm{Q}|\big]^2$, reflecting the isotropic edge positions shown in Fig.~\ref{fig.9}(a3). Then, the $q$-integration in the pseudogap parameter $\Delta_{\rm pg}$ in Eq.~(\ref{eq.PG.gas}) always diverges as far as $\bm{Q}\neq 0$, even in three dimensions. This means that the divergence of the correction term $N_{{\rm FL}, \sigma}$ in the NNSR number equation (\ref{eq.filling.NNSR.gas}). Because of this singularity, the NNSR coupled equations (\ref{eq.NThouless.gas}) and (\ref{eq.filling.NNSR.gas}) are never satisfied simultaneously, which prohibits the NFFLO phase transition.
\par
We note that the essence of the stabilization mechanism of the NFFLO state is, strictly speaking, not the detailed lattice potential itself, but the resulting anisotropy of the `Fermi surface' edges shown in Fig.~\ref{fig.9}(b3). Indeed, as seen in the left panels in Fig.~\ref{fig.10}, when one deforms the shape of these edges to be more spherical by increasing the value of the next nearest neighbor hopping $t'$ (see Fig.~\ref{fig.11}), the NFFLO region obtained in the NNSR theory shrinks. Since the NMF result is not sensitive to $t'$ (see the right panels in Fig.~\ref{fig.10}), the suppression of the NFFLO region seen in Fig.~\ref{fig.10}(a1) is found to be due to stronger NFFLO pairing fluctuations by more spherical `Fermi surface' edges. 
\par
\par
\section{Effects of spin imbalance: Relation to thermal equilibrium FFLO state}
\par
We have shown in Sec.~III and in our recent paper \cite{Kawamura2022} that the removal of the spatial isotropy of a Fermi gas by a three-dimensional cubic optical lattice is a promising route to stabilize both the thermal equilibrium and non-equilibrium FFLO states against pairing fluctuations. In this section, we examine how these FFLO states are related to each other, by considering a spin-{\it imbalanced} driven-dissipative lattice Fermi gas. As shown in Fig.~\ref{fig.8}, once the NFFLO state is stabilized by the optical lattice, the essential behavior of the phase transition temperature $T_{\rm env}^{\rm c}$ can be captured by the simpler NMF theory at $n=0.3$. Keeping this in mind, in this section, we treat the spin-imbalanced system at this filling fraction within the NMF scheme.
\par
\begin{figure}[tb]
\centering
\includegraphics[width=8cm]{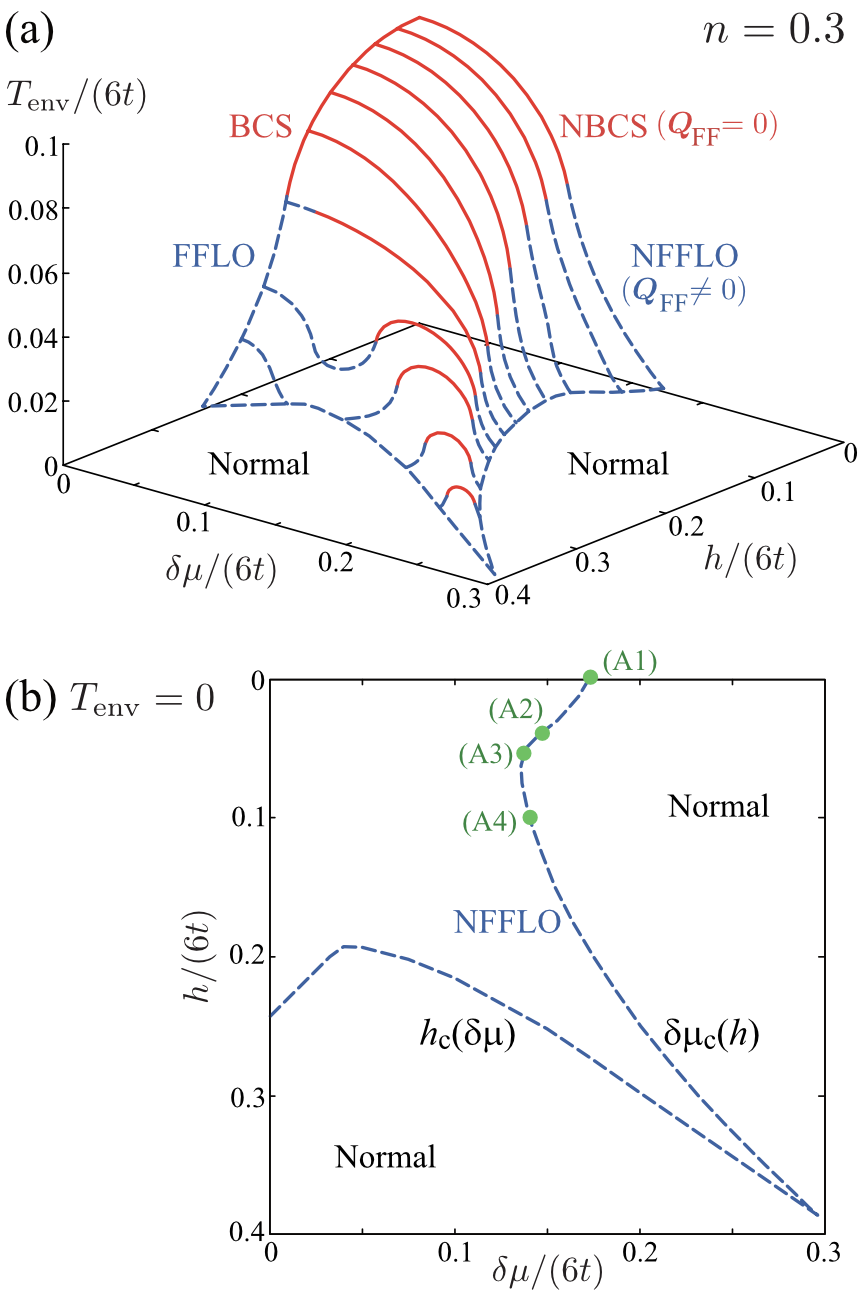}
\caption{(a) Phase diagram of a driven-dissipative lattice Fermi gas, with respect to the environmental temperature $T_{\rm env}$, chemical potential bias $\delta\mu$, and fictitious magnetic field $h$. The solid (dashed) line denotes the NBCS (NFFLO) phase transition temperature $T_{\rm env}^{\rm c}$. $\delta\mu_{\rm c}(h)$ and $h_{\rm c}(\delta\mu)$ are, respectively, the critical chemical potential bias and the critical magnetic field, above which the superfluid phase vanishes. The system is in the thermal equilibrium state at $\delta\mu=0$, where the thermal equilibrium BCS and FFLO states are realized, depending on the magnitude of $h$. (b) The phase diagram at $T_{\rm env}=0$. We set $n=0.3$, $t'=0$, and $\gamma \to +0$, and the NMF theory is used.}
\label{fig.12}
\end{figure}
\par
\begin{figure}[tb]
\centering
\includegraphics[width=8cm]{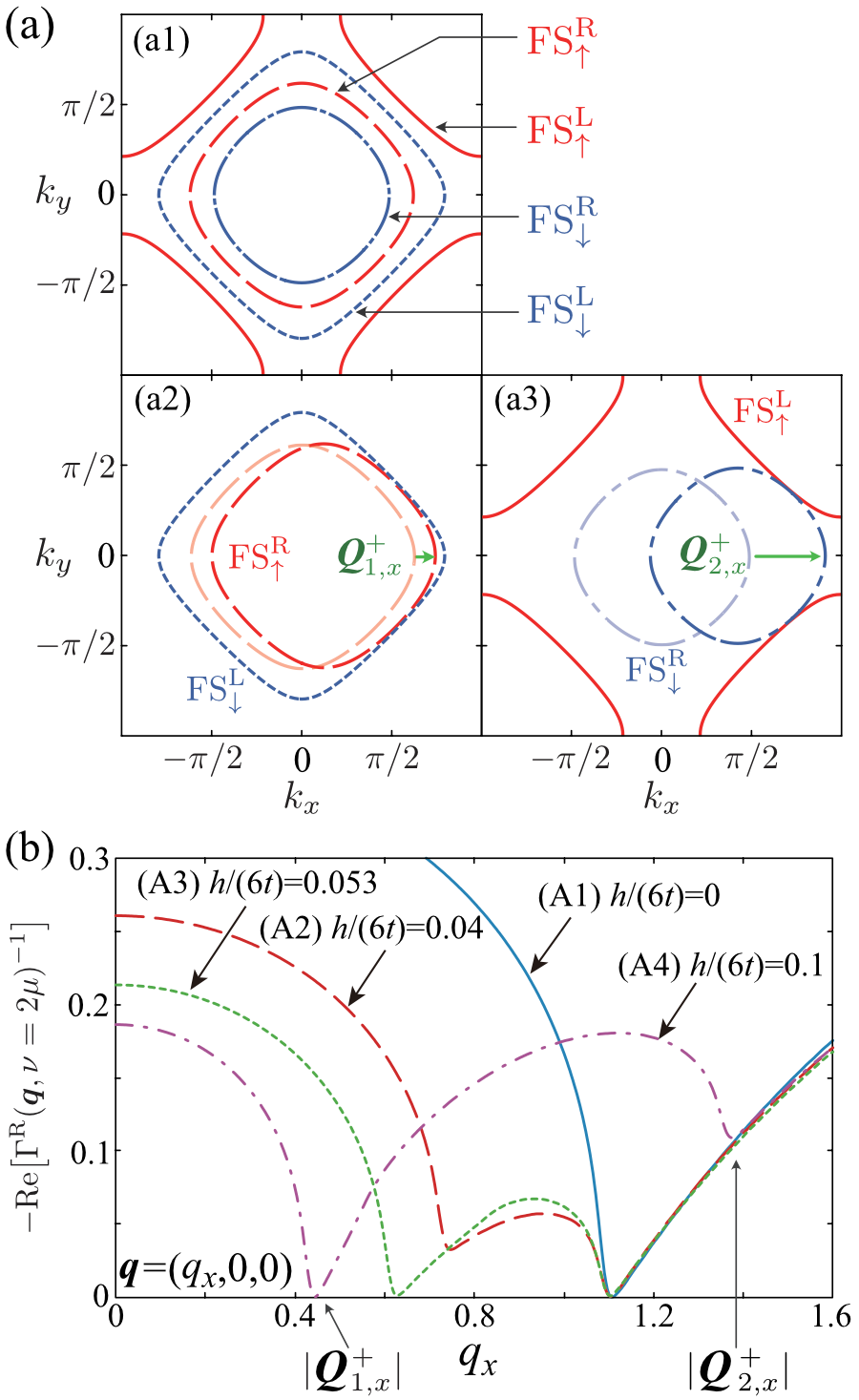}
\caption{(a) (a1) Positions of four edges imprinted on the momentum distribution of Fermi atoms: ${\rm FS}_{\up}^{{\rm L}}$ (solid line), ${\rm FS}_{\up}^{{\rm R}}$ (dashed line), ${\rm FS}_{\down}^{{\rm L}}$ (dotted line), and ${\rm FS}_{\down}^{{\rm R}}$ (dashed-dotted line), at the phase boundary (A4) in Fig.~\ref{fig.12}(b). These lines are obtained from Eqs.~(\ref{eq.eff.kR}) and (\ref{eq.eff.kL}) at $k_z=0$. (a2) Nesting vector $\bm{Q}_{1,x}^+$ between the Fermi surfaces ${\rm FS}_{\up}^{{\rm R}}$ and ${\rm FS}_{\down}^{{\rm L}}$. (a3) Nesting vector $\bm{Q}_{2,x}^+$ between the Fermi surfaces ${\rm FS}_{\down}^{\rm R}$ and ${\rm FS}_{\up}^{\rm L}$. Because of the four-fold symmetry of the background optical lattice, physically equivalent nesting vectors to $\bm{Q}_{1,x}^+$ and $\bm{Q}_{2,x}^+$ also exist in the $-x$-direction, as well as the $\pm y$- and $\pm z$-directions. (b) Inverse retarded particle-particle scattering matrix $-{\rm Re}\big[\Gamma^{\rm R}(\bm{q}=(q_x,0,0), \nu=2\mu)^{-1}\big]$, as a function of $q_x$. Each result is at the phase boundary (A1)-(A4) in Fig.~\ref{fig.12}(b). When $h\ne 0$, $-{\rm Re}\big[\Gamma^{\rm R}(\bm{q},\nu=2\mu)^{-1}\big]$ has two minima at the nesting vectors. As an example, we show the positions of $|\bm{Q}^{+}_{1,x}|$ and $|\bm{Q}^{+}_{2,x}|$ ($>|\bm{Q}^{+}_{1,x}|$) in panel (b), where $\bm{Q}^{+}_{1,x}$ and $\bm{Q}^{+}_{2,x}$ are given in panels (a2) and (a3), respectively.
}
\label{fig.13}
\end{figure}

Figure~\ref{fig.12}(a) shows the phase diagram of a driven-dissipative lattice Fermi gas in the non-equilibrium steady state, with respect to the environmental temperature $T_{\rm env}$, the chemical potential bias $\delta\mu$, and the fictitious magnetic field $h$ to adjust the spin imbalance of the main system. In this figure, the $T_{\rm env}$-$\delta\mu$ plane at $h=0$ describes the spin-{\it balanced} non-equilibrium steady state discussed in Sec.~III, where the two Fermi surface edges produced by the two reservoirs lead to the NFFLO phase transition in the region of large chemical potential bias $\delta\mu$. On the other hand, the $T_{\rm env}$-$h$ plane at $\delta\mu=0$ corresponds to the spin-{\it imbalanced} thermal equilibrium state, where the Zeeman splitting of $\uparrow$-spin and $\downarrow$-spin Fermi surfaces brings about the ordinary FFLO phase transition in the region of high magnetic field $h$ \cite{Fulde1964, Larkin1964, Takada1969, Matsuda2007}. Except for these limits, the system with $\delta\mu\ne 0$ and $h\ne 0$ has four Fermi surface like edges in the momentum distribution of Fermi atoms, as illustrated in Fig.~\ref{fig.1}(c). To be more correct, for example, the two edges at ${\rm FS}^{\rm L}$ and ${\rm FS}^{\rm R}$ in Fig.~\ref{fig.11} ($\delta\mu\ne 0$ and $h=0$), respectively, split into ${\rm FS}_\uparrow^{\rm L}$ and ${\rm FS}_\downarrow^{\rm L}$, and ${\rm FS}_\uparrow^{\rm R}$ and ${\rm FS}_\downarrow^{\rm R}$, as shown in Fig.~\ref{fig.13}(a1). In what follows, we simply call these four edges Fermi surfaces, unless any confusion may occur.
\par
\par
We first focus on the phase diagram at $T_{\rm env}=0$, which is explicitly shown in Fig.~\ref{fig.12}(b). To see the role of the four Fermi surfaces ${\rm FS}_{\sigma=\up,\down}^{\rm L,R}$, we plot in Fig.~\ref{fig.13}(b) the inverse $\Gamma^{\rm R}(\bm{q}=(q_x,0,0),\nu=2\mu)^{-1}$ of the retarded particle-particle scattering matrix at the phase boundaries (A1)-(A4) in Fig.~\ref{fig.12}(b) \cite{note.QFF}. While $-{\rm Re}\big[\Gamma^{\rm R}(\bm{q}, \nu=2\mu)^{-1}\big]$ has a single minimum in the spin-balanced case ($h=0$), it has two dips in the presence of spin imbalance ($h\neq 0$), which physically means the enhancement of pairing fluctuations around these dip momenta. (The same enhancement can also be seen in the $-q_x$ direction, as well as $\pm q_y$ and $\pm q_z$ directions because of the four-fold rotational symmetry of the cubic lattice.) 
\par
We point out that these enhancements of pairing fluctuations around the dip momenta are directly related to the nesting property of the four Fermi surfaces ${\rm FS}_{\sigma={\up,\down}}^{\alpha={\rm L,R}}$ in Fig.~\ref{fig.13}(a1). 
As an example, we show in Fig.~\ref{fig.13}(a2) the nesting vector $\bm{Q}^{+}_{1, x}$ between the Fermi surfaces ${\rm FS}_\down^{\rm L}$ and ${\rm FS}_\up^{\rm R}$ when $h/(6t)=0.1$. (If we translate ${\rm FS}_\up^{\rm R}$ by the momentum $\bm{Q}^{+}_{1, x}$, ${\rm FS}_\up^{\rm R}$ overlaps with ${\rm FS}_\down^{\rm L}$.) We see in Fig.~\ref{fig.13}(b) that the nesting vector $\bm{Q}^{+}_{1, x}$ between ${\rm FS}_\up^{\rm R}$ and ${\rm FS}_\down^{\rm L}$  just give the smaller dip momentum of $-{\rm Re}\big[\Gamma^{\rm R}(\bm{q}, \nu=2\mu)^{-1}\big]$. That is, strong pairing fluctuations around $\bm{Q}^{+}_{1, x}$ are associated with FFLO-type Cooper pairings between Fermi atoms near ${\rm FS}_\up^{\rm R}$ and ${\rm FS}_\down^{\rm L}$. In the same manner, the larger dip momentum seen in Fig.~\ref{fig.13}(b) also equals another nesting vector $\bm{Q}^{+}_{2, x}$ between the ${\rm FS}_\down^{\rm R}$ and ${\rm FS}_\up^{\rm L}$ shown in Fig.~\ref{fig.13}(a3). Thus, strong pairing fluctuations around $\bm{Q}^{+}_{2, x}$ are found to be associated with Cooper pairings between Fermi atoms near ${\rm FS}_\up^{\rm L}$ and ${\rm FS}_\down^{\rm R}$. We briefly note that the Fermi surfaces ${\rm FS}^{\alpha}_\up$ and ${\rm FS}^{\alpha}_\down$ coincide with each other in the spin-balanced case. Pairing fluctuations around $\bm{Q}^{+}_{1, x}$ and $\bm{Q}^{+}_{2, x}$ are then degenerate, so that $-{\rm Re}\big[\Gamma^{\rm R}(\bm{q}, \nu=2\mu)^{-1}\big]$ has a single minimum when $h/(6t)=0$, as shown in Fig.~\ref{fig.13}(b).
\par
Noting that the $T_{\rm env}^{\rm c}$-equation ({\ref{eq.NThouless}) is equivalent to the pole condition for $\Gamma^{\rm R}(\bm{q}, \nu=2\mu)$ (KM theory), we find from Fig.~\ref{fig.13}(b) that, when $h/(6t)<0.053$, NFFLO superfluid phase transition is dominated by the Cooper-pair formation between ${\rm FS}^{\rm L}_\up$ and ${\rm FS}^{\rm R}_\down$ in Fig.~\ref{fig.13}(a). When $h/(6t)> 0.053$, on the other hand, the retarded particle-particle scattering matrix develops a pole at $\bm{q}=\bm{Q}^+_{1, x}$, which means that ${\rm FS}^{\rm R}_\up$ and ${\rm FS}^{\rm L}_\down$ trigger the NFFLO superfluid instability, instead of ${\rm FS}^{\rm L}_\up$ and ${\rm FS}^{\rm R}_\down$.
\par
This switching of the Fermi surfaces that dominantly contribute to the NFFLO superfluid instability [which occurs at $h/(6t)=0.053$], is the key to understanding the non-monotonic behavior of the critical chemical potential bias $\delta\mu_{\rm c}(h)$ (above which the NFFLO state no longer exists) as a function of magnetic field $h$ [see the phase boundary in Fig.~\ref{fig.12}(b)]: When $h/(6t)< 0.053$, the `Fermi momenta' ${\bm k}_{{\rm F}\up}^{\rm L}$ and ${\bm k}_{{\rm F}\down}^{\rm R}$ of the Fermi surfaces ${\rm FS}^{\rm L}_\up$ and ${\rm FS}^{\rm R}_\down$, which dominantly contribute to the NFFLO Cooper-pair formation, are determined by, respectively,
\begin{align}
&\tilde{\ep}_{\bm{k}^{\rm L}_{{\rm F}\up}, \up} = \mu+h +\delta \mu \equiv\mu +\delta\mu_{\rm eff}
,\\
&\tilde{\ep}_{\bm{k}^{\rm R}_{{\rm F}\down}, \down} = \mu-h-\delta \mu \equiv \mu -\delta\mu_{\rm eff}.
\end{align}
Here, ${\tilde \varepsilon}_{{\bm k},\sigma}$ is given in Eq.~(\ref{eq.tild.ep}). Because the mismatch of the two Fermi surfaces is tuned by adjusting $\delta\mu_{\rm eff} =\delta\mu +h$, and the system near $h=0$ is expected to experience the NFFLO instability when $\delta\mu_{\rm eff} \sim \delta\mu_{\rm c}(h=0)$, one finds that $\delta\mu_{\rm c}(h)$ decreases with increasing $h$, which explains the behavior of $\delta\mu_{\rm c}(h)$ seen in the low magnetic field regime in Fig.~\ref{fig.12}(b).
\par
On the other hand, when $h/(6t)>0.053$, the Fermi momenta ${\bm k}_{{\rm F}\down}^{\rm L}$ and ${\bm k}_{{\rm F}\up}^{\rm R}$ of the Fermi surfaces ${\rm FS}^{\rm L}_\down$ and ${\rm FS}^{\rm R}_\up$, that dominantly contribute to the NFFLO phase transition, are determined by, respectively,
\begin{align}
&\tilde{\ep}_{\bm{k}^{\rm L}_{{\rm F}\down}, \down} = \mu -h +\delta \mu \equiv \mu +\delta\mu_{\rm eff},
\\
&\tilde{\ep}_{\bm{k}^{\rm R}_{{\rm F}\up}, \up} = \mu+h -\delta \mu \equiv \mu -\delta\mu_{\rm eff}.
\end{align}
In contrast to the low field case [$t/(6t)<0.053$], $\delta\mu_{\rm eff}=\delta\mu -h$ determines the mismatch of the Fermi surfaces. Thus, simply assuming that the NFFLO instability occurs when $\delta\mu_{\rm eff}\sim\delta\mu_{\rm c}(0)$, one finds that $\delta\mu_{\rm c}(h)$ increases with increasing $h$ in this high magnetic field regime, which is consistent with the behavior of $\delta\mu_{\rm c}(h>0.053)$ seen in Fig.~\ref{fig.12}(b).
\par
We briefly note that the non-monotonic behavior of the critical magnetic field $h_{\rm c}(\delta\mu)$ seen in Fig.~\ref{fig.12}(b) can also be explained in the same manner: As one increases $\delta\mu$ from zero, the Fermi surfaces ${\rm FS}^{\rm L}_\up$ and ${\rm FS}^{\rm R}_\down$ first dominantly contribute to the NFFLO superfluid phase transition (although we do not explicitly show the result corresponding to Fig.~\ref{fig.13}(b) here), giving the decrease of $h_{\rm c}(\delta\mu)$ with increasing $\delta\mu$, as in the case of $\delta\mu_{\rm c}(h\sim 0)$. However, once the dominant Fermi surfaces switch to ${\rm FS}^{\rm L}_\down$ and ${\rm FS}^{\rm R}_\up$, $h_{\rm c}(\delta\mu)$ increases with increasing $\delta\mu$.
\par
\begin{figure}[tb]
\centering
\includegraphics[width=8cm]{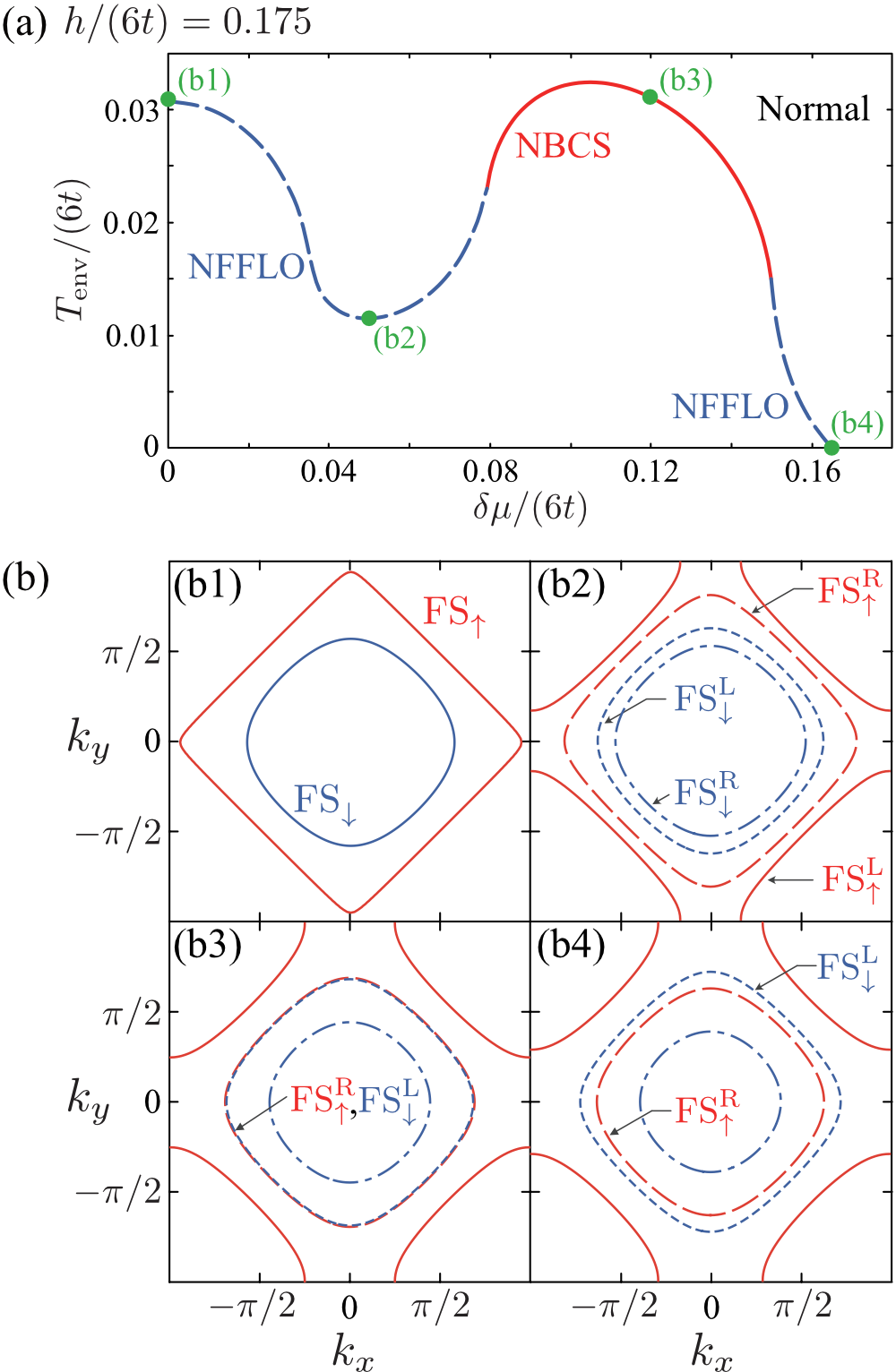}
\caption{(a) Calculated $T_{\rm env}^{\rm c}$ when $h/(6t)=0.175$, as a function of the chemical potential bias $\delta\mu$. (b) Fermi surfaces ${\rm FS}_{\sigma=\up,\down}^{\alpha={\rm L}, {\rm R}}$ ($k_z=0$) at (b1)-(b4) in panel (a). In panel (b1), because $\delta\mu=0$, the two Fermi surfaces ${\rm FS}_\up={\rm FS}_\uparrow^{\rm L,R}$ and ${\rm FS}_\down={\rm FS}_\down^{\rm L,R}$ only exist.}
\label{fig.14}
\end{figure}
\par
We next discuss the $\delta\mu$- and $h$-dependence of the phase transition temperature $T_{\rm env}^{\rm c}$ in Fig.~\ref{fig.12}(a). When the system is out of equilibrium by introducing the chemical potential bias ($\delta\mu>0$), the resulting two-edge structure of the momentum distribution of Fermi atoms works like the thermal broadening of Fermi surfaces, which suppresses the phase transition temperature. Because of this, we see in Fig.~\ref{fig.12}(a) that $T_{\rm env}^{\rm c}$ initially decreases with increasing the chemical potential bias $\delta\mu$.
\par
However, in the high magnetic field regime [$h/(6t) \gtrsim 0.15$], $T_{\rm env}^{\rm c}$ exhibits non-monotonic $\delta\mu$-dependence, as explicitly shown in Fig.~\ref{fig.14}(a). We also see from Fig.~\ref{fig.14}(a) that although the FFLO state appears in the thermal equilibrium state ($\delta\mu=0$), the NBCS state with zero center-of-mass momentum of Cooper pairs appears, when $0.08 \lesssim \delta\mu/(6t) \lesssim 0.15$. To understand the reason for these, we point out that, at (b3) in Fig.~\ref{fig.14}(a), among the four Fermi surfaces, ${\rm FS}_\up^{\rm R}$ and ${\rm FS}_\down^{\rm L}$ are almost degenerate, as shown in Fig.~\ref{fig.14}(b3). Because of this, Cooper pairs are dominantly formed between these two (nearly) degenerate Fermi surfaces, leading to the NBCS superfluid phase transition around (b3) in Fig.~\ref{fig.14}(a).
\par
When one further increases $\delta\mu$ from (b3), the degeneracy of the two Fermi surfaces is lifted, and ${\rm FS}_\down^{\rm L}$ becomes larger than ${\rm FS}_\up^{\rm R}$, as shown in Fig.~\ref{fig.14}(b4). This situation is very similar to the thermal-equilibrium case under an external magnetic field (where $\up$-spin Fermi surface becomes larger than the $\down$-spin Fermi surface). Indeed, Fig.~\ref{fig.14}(a) shows that the superfluid phase transition changes from NBCS to NFFLO at $\delta\mu/(6t)\simeq 0.15$, as in the thermal equilibrium case where the FFLO state appears under a high magnetic field.
\par
When one {\it decreases} $\delta\mu$ from (b3), the degeneracy of the two Fermi surfaces is again lifted, but now ${\rm FS}_\down^{\rm L}$ becomes {\it smaller} than ${\rm FS}_\up^{\rm R}$, as shown in Fig.~\ref{fig.14}(b2). Apart from this difference, the situation is again similar to the thermal equilibrium case under an external magnetic field. Thus, as one decreases $\delta\mu$ from (b3), the NBCS phase transition changes to NFFLO phase transition at $\delta\mu/(6t)\simeq 0.08$, as seen in Fig.~\ref{fig.14}(a). We briefly note that, when the $\delta\mu$ vanishes, the Fermi surfaces ${\rm FS}_\sigma^{\rm L}$ and ${\rm FS}_\sigma^{\rm R}$ are degenerate to each other, so that the Zeeman-split {\it two} Fermi surfaces shown in Fig.~\ref{fig.14}(b1) are restored.
\par
\begin{figure}[tb]
\centering
\includegraphics[width=7.8cm]{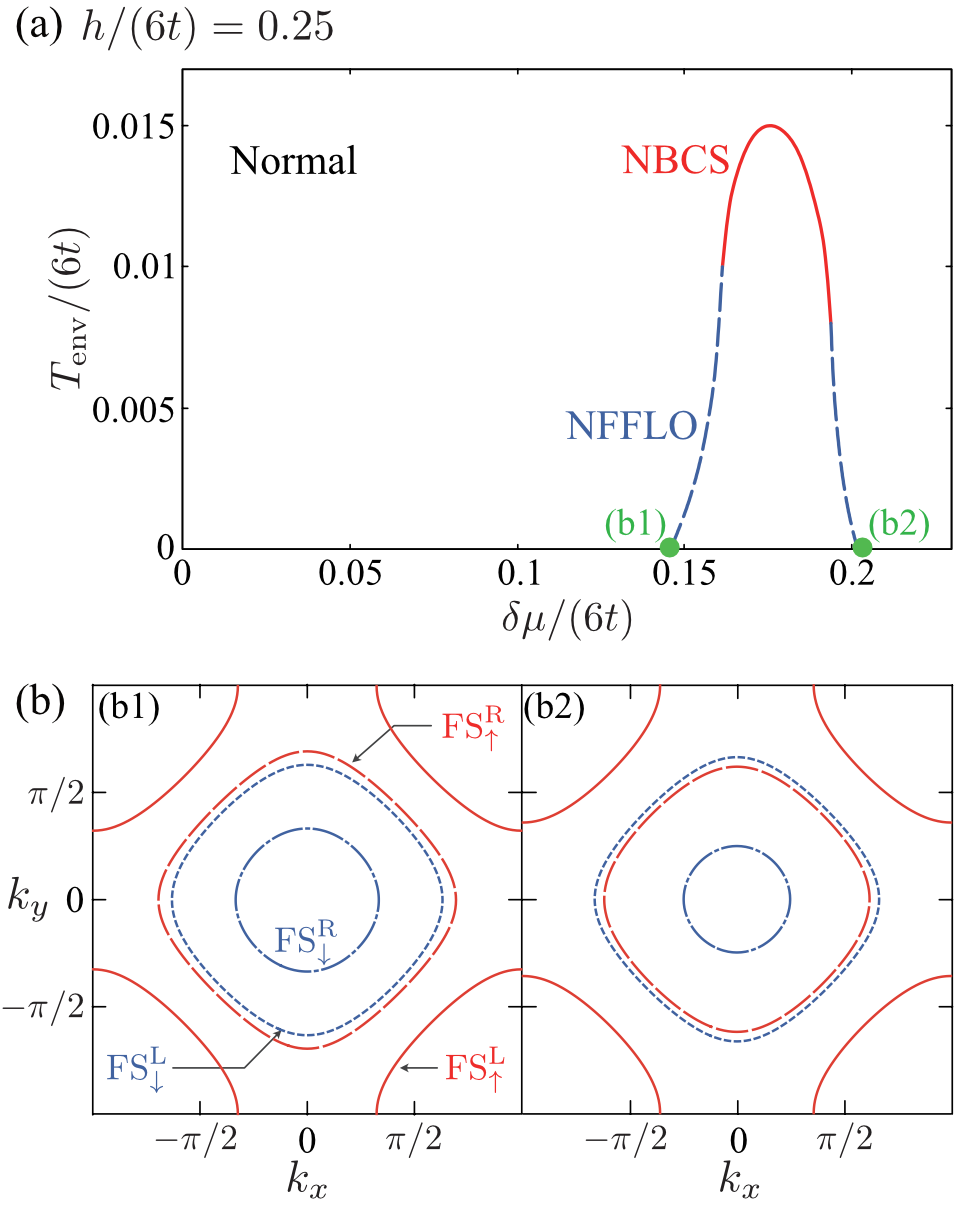}
\caption{(a) Calculated $T_{\rm env}^{\rm c}$ as a function of $\delta\mu$ in the high magnetic field regime [$h/(6t)=0.25$], where the superfluid phase no longer exists in the thermal equilibrium state ($\delta\mu=0$). (b) Fermi surfaces ${\rm FS}_{\sigma=\up,\down}^{\alpha={\rm L}, {\rm R}}$ ($k_z=0$) at (b1) and (b2) in panel (a).}
\label{fig.15}
\end{figure}
\par
We emphasize that Cooper pairings between ${\rm FS}_\up^{\rm R}$ and ${\rm FS}_\down^{\rm L}$ enable a superfluid state even in a high magnetic field where the FFLO state can not be realized in the thermal equilibrium state. To explicitly demonstrate this, we show in Fig.~\ref{fig.15}(a) the calculated $T_{\rm env}^{\rm c}$, when $h/(6t)=0.25~[>h_{\rm c}(\delta\mu=0)]$. Under this high magnetic field, the thermal equilibrium state ($\delta\mu=0$) is in the normal phase down to $T_{\rm env}=0$, because the misalignment between the Zeeman-splitting between the $\up$-spin and $\down$-spin Fermi surfaces is too large to form Cooper pairs there. However, as $\delta\mu$ increases and the main system is driven out of equilibrium, among the four Fermi surfaces, ${\rm FS}_\down^{\rm L}$ and ${\rm FS}_\up^{\rm R}$ become close to each other [see Fig.~\ref{fig.15}(b1)], which enables the NFFLO phase transition, as shown in Fig.~\ref{fig.15}(a). As $\delta\mu$ increases, these two Fermi surfaces become almost degenerate, so that the NFFLO phase transition changes to the NBCS one. This degeneracy is again lifted with further increasing $\delta\mu$, and ${\rm FS}_\down^{\rm L}$ eventually becomes larger than ${\rm FS}_\up^{\rm R}$ [see Fig.\ref{fig.15}(b2)]. Then, the system again experiences the NFFLO phase transition, as shown in Fig.~\ref{fig.15}(a).
\par
\section{Summary}
\par
To summarize, we have studied non-equilibrium superfluid phase transitions in a driven-dissipative lattice Fermi gas coupled with two reservoirs. To include non-equilibrium pairing fluctuations, we extended the thermal-equilibrium strong-coupling theory developed by Nozi\`eres and Schmitt-Rink to the non-equilibrium steady state, by employing the Keldysh Green's function technique. Using this, we showed that a two-edge structure of the Fermi momentum distribution, which is produced by the chemical potential difference between the two reservoirs, makes the system similar to conduction electrons in metals under an external magnetic field. As a result, non-equilibrium Fulde-Ferrell-Larkin-Ovchinnikov (NFFLO) superfluid phase transition was found to occur without spin imbalance. Since this unconventional Fermi superfluid is known to be unstable against pairing fluctuations in a spatially isotropic gas, we pointed out that the removal of the spatial isotropy by the optical lattice is essentially important for the stabilization of the NFFLO state. We also confirmed that, once the NFFLO state is stabilized by the optical lattice, the essential behavior of this superfluid phase transition can be captured within the non-equilibrium mean-field BCS theory, at least when the filling fraction equals $n=0.3$.
\par
We have also examined the case when the system is accompanied by spin imbalance. Within the framework of the non-equilibrium mean-field theory at $n=0.3$, we identified the region where the NFFLO and the thermal equilibrium FFLO states appear, in the phase diagram with respect to the environmental temperature $T_{\rm env}$, the chemical bias $\delta\mu$, and the fictitious magnetic field $h$. 
\par
When $\delta\mu\ne0$ and $h\ne 0$, the two-edge structure imprinted on the momentum distribution of Fermi atoms and the Zeeman splitting of $\up$-spin and $\down$-spin Fermi surfaces coexist, so that the system looks as if it has four different Fermi surfaces. We clarified that the non-monotonic behavior of the critical chemical potential bias $\delta\mu_{\rm c}(h)$ as a function of $h$, as well as the critical magnetic field $h_{\rm c}(\delta\mu)$ as a function of $\delta\mu$, can consistently be explained by using the existence of these four split Fermi surfaces.
\par
Regarding the experimental approach to such multiple Fermi surface effects, a voltage-biased superconducting wire and thin film under an external magnetic field would be promising candidates. Indeed, Refs.\cite{Keizer2006, Vercruyssen2012, Seja2021} reported that the momentum distribution of conduction electrons in such systems is highly out of equilibrium, and exhibits a two-step structure. Thus, by applying an external magnetic field to such systems, not only the non-equilibrium splitting, but also the Zeeman splitting of Fermi surfaces would occur. Then, the resulting four Fermi surfaces would lead to exotic superconducting phase transitions, as discussed in this paper. Indeed, it has been proposed that a voltage-biased superconductor may be used to recover the superconducting state in a high magnetic field beyond the Chandrasekhar-Clogston limit \cite{Ouassou2018}, just as in Fig.~\ref{fig.15}(a); however, the possibility of the NFFLO phase transition is not discussed in Ref. \cite{Ouassou2018}.
\par
In this paper, we have focused on the superfluid phase transition temperature $T_{\rm env}^{\rm c}$. Extension of this work to the superfluid phase below $T_{\rm env}^{\rm c}$ to clarify how the multiple Fermi surfaces affect superfluid properties is an important future challenge. In addition, the driven-dissipative ultracold Fermi gas system is known to exhibit bi-stability, where two stable states are obtained for the same environmental parameters \cite{Kawamura2022}. Thus, it would also be a crucial future problem to identify the region where this phenomenon occurs, in the phase diagram of the driven-dissipative spin-imbalanced lattice Fermi gas. Since the realization of unconventional Fermi superfluids is one of the most exciting challenges in cold atom physics, our results would be helpful for the study toward the realization of the FFLO superfluid Fermi gas. In addition, the combination of the Zeeman splitting and the non-equilibrium splitting of Fermi surfaces discussed in this paper can be considered, not only in ultracold Fermi gases, but also in other systems, such as a voltage-biased metallic superconductor under an external magnetic field. Thus, our results would also widely contribute to the further development of non-equilibrium condensed matter physics.
\par
\begin{acknowledgments}
\par
T.K. was supported by MEXT and JSPS KAKENHI Grant-in-Aid for JSPS fellows Grant No.JP21J22452. D.K. was supported by JST FOREST (Grant No. JPMJFR202T). Y.O. was supported by a Grant-in-aid for Scientific Research from MEXT and JSPS in Japan (No.JP18K11345, No.JP18H05406, No.JP19K03689, and No.JP22K03486).
\end{acknowledgments}
\appendix
\begin{widetext}
\section{Computational details of filling fraction $n_\sigma$ in Eq.~(\ref{eq.filling.NNSR})}
\par
In this paper, to numerically evaluate the NNSR filling fraction $n_\sigma$ in Eq.~(\ref{eq.filling.NNSR}), we apply the Fourier transform technique \cite{Haussmann1994, Serene1991, Haussmann2007} to the $\bm{k}$-summations in Eqs.~(\ref{eq.PiR}), (\ref{eq.PiK}), (\ref{eq.selfR.NNSR}), and (\ref{eq.selfK.NNSR}). Real space expressions for the pair correlation functions in Eqs.~(\ref{eq.PiR}) and (\ref{eq.PiK}), as well as the NNSR self-energies in Eqs.~\eqref{eq.selfR.NNSR} and \eqref{eq.selfK.NNSR}, are given by, respectively,
\begin{align}
\Pi^{\rm R}(\bm{r}, \nu)
&= 
\big[\Pi^{\rm A}(\bm{r}, \nu)\big]^*
\notag\\[4pt]
&=
\frac{i}{2}\int_{-\infty}^\infty \frac{d\omega}{2\pi}
\Big[G^{\rm R}_{{\rm NMF}, \up}(\bm{r}, \omega +\nu) G^{\rm K}_{{\rm NMF}, \down}(\bm{r}, -\omega) +
G^{\rm K}_{{\rm NMF}, \up}(\bm{r}, \omega +\nu) G^{\rm R}_{{\rm NMF}, \down}(\bm{r}, -\omega)\Big]
\label{eq.A1}
,\\[6pt]
\Pi^{\rm K}(\bm{r}, \nu)&=
\frac{i}{2}\int_{-\infty}^\infty \frac{d\omega}{2\pi}
\Big[G^{\rm R}_{{\rm NMF}, \up}(\bm{r}, \omega +\nu) G^{\rm R}_{{\rm NMF}, \down}(\bm{r}, -\omega) +
G^{\rm A}_{{\rm NMF}, \up}(\bm{r}, \omega +\nu) G^{\rm A}_{{\rm NMF}, \down}(\bm{r}, -\omega) 
\notag\\
&\hspace{1.8cm}+
G^{\rm K}_{{\rm NMF}, \up}(\bm{r}, \omega +\nu) G^{\rm K}_{{\rm NMF}, \down}(\bm{r}, -\omega) \Big],
\\[4pt]
\Sigma^{\rm R}_{{\rm NNSR},\sigma}(\bm{r}, \omega) 
&=
\big[\Sigma^{\rm A}_{{\rm NNSR},\sigma}(\bm{r}, \omega)\big]^*
\notag\\[4pt]
&=
-\frac{i}{2} \int_{-\infty}^\infty \frac{d\nu}{2\pi}\Big[
\Gamma^{\rm R}(\bm{r}, \nu) G^{\rm K}_{{\rm NMF}, -\sigma}(-\bm{r}, \nu -\omega) +
\Gamma^{\rm K}(\bm{r}, \nu) G^{\rm A}_{{\rm NMF}, -\sigma}(-\bm{r}, \nu -\omega)\Big]
,\\[6pt]
\Sigma^{\rm K}_{{\rm NNSR},\sigma}(\bm{r}, \omega) 
&=
-\frac{i}{2}\int_{-\infty}^\infty \frac{d\nu}{2\pi}\Big[
\Gamma^{\rm A}(\bm{r}, \nu) G^{\rm R}_{{\rm NMF}, -\sigma}(-\bm{r}, \nu -\omega) +
\Gamma^{\rm R}(\bm{r}, \nu) G^{\rm A}_{{\rm NMF}, -\sigma}(-\bm{r}, \nu -\omega)
\notag\\
&\hspace{2cm}+
\Gamma^{\rm K}(\bm{q}, \nu) G^{\rm K}_{{\rm NMF}, -\sigma}(-\bm{r}, \nu -\omega)
\Big].
\label{eq.A4}
\end{align} 
\end{widetext}
Using these, one can avoid the heavy ${\bm k}$-summation. To take advantage of this benefit in real space, we employ the fast Fourier transformation (FFT) method to execute the following Fourier transformation:
\begin{equation}
\hat{G}_{{\rm NMF}, \sigma} (\bm{r}, \omega) = \sum_{\bm{k}} \hat{G}_{{\rm NMF}, \sigma} (\bm{k}, \omega) e^{i\bm{k}\cdot\bm{r}}. 
\label{eq.FFT.G}
\end{equation}
The combination of the FFT method and the real space expressions in Eqs.~\eqref{eq.A1}-\eqref{eq.A4} significantly reduces the computational cost compared to the direct evaluation of the $\bm{k}$-summations.
\par
When the damping rate $\gamma$ becomes small, the NMF Green's function $\hat{G}_{{\rm NMF}, \sigma} (\bm{k}, \omega)$ in Eq.~(\ref{eq.GNMF2}) has a very sharp peak in $\bm{k}$-space. This requires a large number of meshes in momentum space, in order to keep the high accuracy of the FFT method. In our computations, we thus have discretized the three-dimensional momentum region $0 \leq k_j \leq \pi$ ($j=x,y,z$) into $64 \times 64 \times 64$ cells in Eq.~(\ref{eq.FFT.G}). To achieve sufficient accuracy with this number of meshes, $\gamma$ needs to be chosen as $\gamma/(6t)\gesim 0.005$.
\par
\begin{widetext}
\par
\section{NMF and NNSR theories in the absence of optical lattice}
\par
We summarize the NMF theory, as well as the NNSR theory, in the absence of an optical lattice. In this case, the main system in Fig.~\ref{fig.2}(a) becomes a spatially isotropic gas with the ordinary kinetic energy of a free particle,
\begin{equation}
\ep_{\bm k}^{\rm free}={\bm{k}^2 \over 2m}.
\end{equation}
The momentum ${\bm k}$ is ${\it not}$ restricted to the first Brillouin zone, which is in contrast to the lattice system. The $s$-wave interaction term in Eq.~(\ref{eq.Hsys}) then involves the ultraviolet divergence, so that, as usual, we measure the interaction strength in terms of the $s$-wave scattering length $a_s$, in order to remove this singularity from the theory \cite{RanderiaBook}. The scattering length $a_s$ is related to the bare interaction $-U$ as,
\begin{equation}
\frac{4\pi a_s}{m} = \frac{-U}{1 -U\sum_{\bm{k}}^{k_{\rm c}} \frac{1}{2\ep^{\rm free}_{\bm{k}}}},
\end{equation}
where $k_{\rm c}$ is a momentum cutoff, which is eventually taken to be infinity. 
\par
A crucial difference from the lattice system is the vanishing Hartree term, because $U\to+0$ in the limit $p_{\rm c}\to\infty$.\cite{Wyk2018, Ohashi2020}. The $T_{\rm env}^{\rm c}$-equation (\ref{eq.NThouless}) in the absence of the optical lattice is then reduced to
\begin{equation}
1=U\gamma\sum_{\bm{k}} \int_{-\infty}^\infty \frac{d\omega}{2\pi}
\frac
{\big[2\omega + \ep^{\rm free}_{\bm{k}+\bm{q}/2} -\ep^{\rm free}_{-\bm{k}+\bm{q}/2} -2h\big]
\left[\tanh\left(\frac{\omega-\delta\mu}{2T^{\rm c}_{\rm env}}\right) +\tanh\left(\frac{\omega-\delta\mu}{2T^{\rm c}_{\rm env}}\right)\right]}
{\big[(\omega +\ep^{\rm free}_{\bm{k}+\bm{q}/2} -\mu_\up)^2 +4\gamma^2 \big]
 \big[(\omega -\ep^{\rm free}_{\bm{k}+\bm{q}/2} +\mu_\down)^2 +4\gamma^2 \big]}.
\label{eq.NThouless.gas}	
\end{equation}
In the same manner, the equation for the filling fraction in Eq.~(\ref{eq.n.NMF}) is replaced by the number equation,
\begin{equation}
N_{{\rm NMF},\sigma}= 
\sum_{\bm{k}} \int_{-\infty}^\infty \frac{d\omega}{2\pi} \frac{4\gamma}{[\omega -\ep^{\rm free}_{\bm{k}}]^2 +4\gamma^2}
\big[f(\omega -\mu_{{\rm L}, \sigma}) +f(\omega -\mu_{{\rm R}, \sigma})\big],
\label{eq.n.NMF.gas}
\end{equation}
where $N_{{\rm NMF},\sigma}$ is the total number of Fermi atoms in the $\sigma$-spin component in the main system.
\par
We next explain the NNSR theory. In the absence of the Hartree term, the Green's function ${\hat G}_{{\rm env},\sigma}$ in Eq.~(\ref{eq.Genv}) equals ${\hat G}_{{\rm NMF},\sigma}$ in Eq.~(\ref{eq.GNMF2}). In the absence of the optical lattice, thus, the pair-correlation functions $\Pi^{\rm R,A,K}({\bm q},\nu)$, as well as the NNSR self-energies $\Sigma_{{\rm NNSR},\sigma}^{\rm R,A,K}({\bm k},\nu)$, can be constructed by using
\begin{equation}
\hat{G}_{{\rm env}, \sigma}(\bm{p}, \omega)=
\begin{pmatrix}
\frac{1}{\omega -\ep^{\rm free}_{\bm{p}} +2i\gamma} &
\frac{-4i\gamma[1 -f(\omega -\mu_{{\rm L},\sigma}) -f(\omega -\mu_{{\rm R},\sigma})]}
{[\omega -\ep^{\rm free}_{\bm{p}}]^2 +4\gamma^2}\\[6pt]
0 &
\frac{1}{\omega -\ep^{\rm free}_{\bm{p}} -2i\gamma}
\end{pmatrix},
\label{eq.Genv.gas}
\end{equation}
The resulting expressions are
\begin{align}
\Pi^{\rm R}(\bm{q}, \nu)
&=
\big[\Pi^{\rm A}(\bm{q}, \nu)\big]^*
\notag\\
&=
\frac{i}{2} \sum_{\bm{p}} \int_{-\infty}^\infty \frac{d\omega}{2\pi}
\Big[G^{\rm R}_{{\rm env}, \up}(\bm{p}+\bm{q}/2, \omega +\nu) G^{\rm K}_{{\rm env}, \down}(-\bm{p}+\bm{q}/2, -\omega) 
\notag\\
&\hspace{4.5cm}+
G^{\rm K}_{{\rm env}, \up}(\bm{p}+\bm{q}/2, \omega +\nu) G^{\rm R}_{{\rm env}, \down}(-\bm{p}+\bm{q}/2, -\omega)\Big]
\label{eq.PiR.gas}
,\\
\Pi^{\rm K}(\bm{q}, \nu)
&=
\frac{i}{2} \sum_{\bm{p}} \int_{-\infty}^\infty \frac{d\omega}{2\pi}
\Big[G^{\rm R}_{{\rm env}, \up}(\bm{p}+\bm{q}/2, \omega +\nu) G^{\rm R}_{{\rm env}, \down}(-\bm{p}+\bm{q}/2, -\omega) 
\notag\\
&\hspace{4.5cm}+
G^{\rm A}_{{\rm env}, \up}(\bm{p}+\bm{q}/2, \omega +\nu) G^{\rm A}_{{\rm env}, \down}(-\bm{p}+\bm{q}/2, -\omega)
\notag\\
&\hspace{4.5cm}+
G^{\rm K}_{{\rm env}, \up}(\bm{p}+\bm{q}/2, \omega +\nu) G^{\rm K}_{{\rm env}, \down}(-\bm{p}+\bm{q}/2, -\omega)\Big],
\label{eq.PiK.gas}\\
\Sigma^{\rm R}_{{\rm NNSR},\sigma}(\bm{k}, \omega) &=
\big[\Sigma^{\rm A}_{{\rm NNSR},\sigma}(\bm{k}, \omega)\big]^*
\notag\\[4pt]
&=
-\frac{i}{2} \sum_{\bm{q}} \int_{-\infty}^\infty \frac{d\nu}{2\pi}\Big[
\Gamma^{\rm R}(\bm{q}, \nu) G^{\rm K}_{{\rm env}, -\sigma}(\bm{q}-\bm{k}, \nu -\omega) +
\Gamma^{\rm K}(\bm{q}, \nu) G^{\rm A}_{{\rm env}, -\sigma}(\bm{q}-\bm{k}, \nu -\omega)
\Big]
\label{eq.selfR.NNSR.gas}
,\\[6pt]
\Sigma^{\rm K}_{{\rm NNSR},\sigma}(\bm{k}, \omega) &=
-\frac{i}{2} \sum_{\bm{q}} \int_{-\infty}^\infty \frac{d\nu}{2\pi}\Big[
\Gamma^{\rm A}(\bm{q}, \nu) G^{\rm R}_{{\rm env}, -\sigma}(\bm{q}-\bm{k}, \nu -\omega)
\notag\\
&\hspace{2.65cm}+
\Gamma^{\rm R}(\bm{q}, \nu) G^{\rm A}_{{\rm env}, -\sigma}(\bm{q}-\bm{k}, \nu -\omega) +
\Gamma^{\rm K}(\bm{q}, \nu) G^{\rm K}_{{\rm env}, -\sigma}(\bm{q}-\bm{k}, \nu -\omega)
\Big].
\label{eq.selfK.NNSR.gas}
\end{align}
Here, the particle-particle scattering matrix $\hat{\Gamma}(\bm{q}, \nu)$ is given in Eq.~(\ref{eq.Tmat}). The NNSR Green's function $\hat{G}_{{\rm NNSR}, \sigma}$ in the absence of the optical lattice is then given by
\begin{equation}
\hat{G}_{{\rm NNSR}, \sigma}(\bm{k}, \omega)=
\hat{G}_{{\rm env}, \sigma}(\bm{k}, \omega) +
\hat{G}_{{\rm env}, \sigma}(\bm{k}, \omega) 
\hat{\Sigma}_{{\rm NNSR}, \sigma}(\bm{k}, \omega)
\hat{G}_{{\rm env}, \sigma}(\bm{k}, \omega).
\end{equation}
Using the Keldysh component of the NNSR Green's function $\hat{G}_{{\rm NNSR}, \sigma}$, we find that the total number $N_\sigma$ of Fermi atoms with $\sigma$-spin in the main system can be written as
\begin{align}
N_\sigma 
&= \frac{i}{2} \sum_{\bm{k}} \int_{-\infty}^\infty \frac{d\omega}{2\pi} 
G^{\rm K}_{{\rm NNSR}, \sigma}(\bm{k}, \omega) -\frac{1}{2}
\notag\\
&= 
N_{{\rm NMF}, \sigma} + \frac{i}{2} 
\sum_{\bm{k}} \int_{-\infty}^\infty \frac{d\omega}{2\pi}
\Big[ 
\hat{G}_{{\rm env}, \sigma}(\bm{k}, \omega) 
\hat{\Sigma}_{{\rm NNSR}, \sigma}(\bm{k}, \omega)
\hat{G}_{{\rm env}, \sigma}(\bm{k}, \omega)
\Big]^{\rm K}
\notag\\
&\equiv 
N_{{\rm NMF}, \sigma} + N_{{\rm FL},\sigma},
\label{eq.filling.NNSR.gas}
\end{align}
where $N_{{\rm NMF},\sigma}$ is given in Eq.~(\ref{eq.n.NMF.gas}). As in the lattice system, we solve the NNSR number equation (\ref{eq.filling.NNSR.gas}), together with the $T_{\rm env}^{\rm c}$-equation (\ref{eq.NThouless.gas}), to self-consistently determine $T_{\rm env}^{\rm c}$, $\mu_\uparrow$, as well as $\mu_\downarrow$.
\par
\par
\section{Destruction of NFFLO long-range order in the absence of optical lattice}
\par
In this appendix, we show that, when the optical lattice is absent, any NFFLO solution with ${\bm Q}_{\rm FF}\ne 0$ cannot simultaneously satisfy the $T_{\rm env}^{\rm c}$-equation (\ref{eq.NThouless.gas}) and the NNSR number equation (\ref{eq.filling.NNSR.gas}), because the fluctuation correction term $N_{{\rm FL}, \sigma}$ involved in the number equation always diverges when the $T_{\rm env}^{\rm c}$-equation is satisfied. 
\par
When the $T_{\rm env}^{\rm c}$-equation (\ref{eq.NThouless.gas}) is satisfied at a parameter set $(T_{\rm env}, \mu, {\bm q})=(T_{\rm env}^{\bm Q}, \mu_{\bm Q}, {\bm Q})$, the particle-particle scattering matrix $\hat{\Gamma}({\bm Q}, \nu=2\mu_{\bm Q})$ also diverges at $T_{\rm env}^{\bm Q}$. Thus, the self-energies in Eqs.~(\ref{eq.selfR.NNSR.gas}) and (\ref{eq.selfK.NNSR.gas}) at $T_{\rm env}^{\bm Q}$ may be approximated to
\begin{align}
\hat{\Sigma}_{{\rm NNSR},\sigma}(\bm{k}, \omega)	
& \simeq
-\Delta_{\rm pg}^2
\begin{pmatrix}
G^{\rm A}_{{\rm env}, -\sigma}(\bm{Q}-\bm{k}, 2\mu -\omega) & 
G^{\rm K}_{{\rm env}, -\sigma}(\bm{Q}-\bm{k}, 2\mu -\omega) \\[4pt]
0 & 
G^{\rm R}_{{\rm env}, -\sigma}(\bm{Q}-\bm{k}, 2\mu -\omega)
\end{pmatrix}
\notag\\[4pt]
&=
\Delta_{\rm pg}^2 \hat{G}^*_{{\rm env}, -\sigma}(\bm{Q}-\bm{k}, 2\mu -\omega).
\label{eq.PG.self.gas}
\end{align}
Here, the pseudogap parameter $\Delta_{\rm pg}^2$ is given in Eq.~(\ref{eq.Delta.PG}). Substituting Eq.~(\ref{eq.PG.self.gas}) into the number equation (\ref{eq.filling.NNSR.gas}), one obtains
\begin{equation}
N_{{\rm FL}, \sigma}\simeq 
\frac{i\Delta_{\rm pg}^2}{2}
\sum_{\bm{p}} \int_{-\infty}^\infty \frac{d\omega}{2\pi}
\Big[ 
\hat{G}_{{\rm env}, \sigma}(\bm{p}, \omega) 
\hat{G}^*_{{\rm env}, -\sigma}(\bm{Q}-\bm{p}, 2\mu -\omega)
\hat{G}_{{\rm env}, \sigma}(\bm{p}, \omega)
\Big]^{\rm K}.
\label{eq.app.NFL.gas}
\end{equation}
The absence of the cubic optical lattice recovers the spatial isotropy of the main system, so that the retarded particle-particle scattering matrix $\Gamma^{\rm R}(\bm{q}, \nu)$ behaves as, around $(\bm{q}, \nu)= (\bm{Q}, 2\mu)$ \cite{Ohashi2002}, 
\begin{equation}
\Gamma^{\rm R}(\bm{q}, \nu) \simeq
\frac{-U}{C\big[|\bm{q}| -|\bm{Q}|\big]^2 -i\lambda \big[\nu -2\mu\big]}.
\label{eq.app.TR.gas}
\end{equation}
Here,
\begin{equation}
C=\frac{U}{2} \left.\frac{\partial^2 \Pi^{\rm R}(\bm{q}, 2\mu_{\bm Q})}{\partial |\bm{q}|^2} \right|_{\bm{q}=\bm{Q}}
\end{equation}
and $\lambda$ is given in Eq.~(\ref{eq.lambda}). In obtaining Eq.~\eqref{eq.app.TR.gas}, we have taken the limit $\gamma \to +0$, for simplicity. Using Eqs.~(\ref{eq.Tmat}) and (\ref{eq.app.TR.gas}), one can evaluate the pseudogap parameter $\Delta_{\rm pg}^2$ in Eq.~(\ref{eq.Delta.PG}) as
\begin{align}
\Delta_{\rm pg}^2 
&=
\frac{i}{2} \sum_{\bm{q}} \int_{-\infty}^\infty \frac{d\nu}{2\pi} 
\left|\Gamma^{\rm R}(\bm{q}, \nu)\right|^2 \Pi^{\rm K}(\bm{q}, \nu)
\notag\\
&\simeq
\frac{iU^2\Pi^{\rm K}(\bm{Q}, 2\mu)}{2} \sum_{\bm{q}} \int_{-\infty}^\infty \frac{d\nu}{2\pi} 
\frac{1}{C^2 \big[|\bm{q}| -|\bm{Q}|\big]^4 +\lambda^2 \big[\nu-2\mu\big]^2}
\notag\\
&=
\frac{iU^2\Pi^{\rm K}(\bm{Q}, 2\mu)}{4 \lambda C} 
\int_0^{q_{\rm c}}\frac{q^2 dq}{2\pi^2} \frac{1}{\big[|\bm{q}| -|\bm{Q}|\big]^2},
\label{eq.PG.gas.appendix}
\end{align}
where $q_{\rm c}$ is a momentum cutoff. Since the momentum integration in Eq.~(\ref{eq.PG.gas.appendix}) always diverges unless $\bm{Q}=0$, the 
gap parameter $\Delta_{\rm pg}^2$, as well as the fluctuation correction $N_{{\rm FL}, \sigma}$ involved in the NNSR number equation (\ref{eq.filling.NNSR.gas}) (which is proportional to $\Delta_{\rm pg}^2$), diverge at $T_{\rm env}^{\bm Q}$. Thus, the $T_{\rm env}^{\rm c}$-equation (\ref{eq.NThouless.gas}) and the number equation (\ref{eq.filling.NNSR.gas}) are incompatible as far as ${\bm Q}_{\rm FF}\ne 0$.
\end{widetext}
\par


\begin{thebibliography}{99}
\bibitem{note.QFF} We have numerically confirmed that the pairing fluctuations characterized by $\bm{q}\propto (1,0,0) $ are always dominant over those by $\bm{q}\propto (1,1,0)$ or $\bm{q}\propto (1,1,1)$.
\end{thebibliography}
\end{document}